\documentclass[12pt, draftclsnofoot, onecolumn]{IEEEtran}
\usepackage{graphicx}
\usepackage{epstopdf} 
\usepackage{verbatim} 
\usepackage{multirow} 
\usepackage{multicol}
\usepackage{amsfonts}
\usepackage{bbm}
\usepackage{subfigure}
\usepackage{colortbl} 
\usepackage{indentfirst}
\usepackage{dcolumn,booktabs}
\usepackage{algorithm}
\usepackage[noend]{algpseudocode}
\usepackage{algorithmicx,algorithm}
\usepackage{setspace}
\usepackage{algpseudocode}
\usepackage{graphics}
\usepackage{epsfig,color}
\usepackage{amsmath}
\usepackage{amssymb}
\ifCLASSINFOpdf

\else
\fi
\newtheorem{Theorem}{Theorem}

\newtheorem{Lemma}{Lemma}

\begin{document}
\title{On Meta Distribution and Local Delay for Cache-Enabled Networks with Random DTX: Analysis and Optimization}

\author{
\IEEEauthorblockN{Le Yang, Fu-Chun Zheng, \emph{Senior Member}, \emph{IEEE}, Yi Zhong, \emph{Member}, \emph{IEEE}, and Shi Jin,  \emph{Senior Member}, \emph{IEEE}\\}
\IEEEauthorblockA{Emails: \{yangle, jinshi\}@seu.edu.cn, fzheng@ieee.org, yzhong@hust.edu.cn}

}

\maketitle

\begin{abstract}
A fine-grained analysis of the cache-enabled networks is crucial for system design. In this paper, we focus on the meta distribution of the signal-to-interference ratio (SIR) for the cache-enabled networks where the locations of the base stations (BSs) are modeled as a Poisson point process (PPP). With the application of the random caching and the random discontinuous transmission (DTX) schemes, we derive the moments of the conditional successful transmission probability (STP), the exact meta distribution and its beta approximation by utilizing stochastic geometry. The closed-form expressions of the mean and variance of the local delay (i.e., the network jitter) are also derived. We then consider the maximization of the mean STP and the minimization of the average system transmission delay by jointly optimizing the caching probability and the BS active probability. Finally, the numerical results demonstrate the superiority of the proposed optimization schemes over the existing caching strategies and reveal the impacts of the key network parameters on the cache-enabled networks in terms of mean STP, STP variance, meta distribution, mean local delay and network jitter.
\end{abstract}

%
\IEEEpeerreviewmaketitle

\section{Introduction}
\subsection{Motivation}
Due to the rapid proliferation of various multi-media applications and smart mobile devices, the mobile data traffic has witnessed an unprecedented growth and imposed heavy burden on the backhaul links. Moreover, only 5$\%$-10$\%$ of all the files are required by the majority of the users \cite{caching-content}. Stimulated by these facts, equipping base stations (BSs) with caches to pre-fetch the popular contents or files during the off-peak time has emerged as a promising approach to deal with the potential bottleneck issue of backhauling.

The performance evaluation of the cache-enabled networks has become an important issue and most existing works utilize the mean successful transmission probability (STP) as a key performance metric. The STP is the probability of the signal-to-interference ratio (SIR) at a typical user exceeding some threshold $\theta$, i.e., $\mathbbm{P}[\text{SIR}>\theta]$, given a spatial realization of BSs and users. The mean STP is achieved by averaging the STP (a random variable) over the BS distribution (e.g., a point process) and the corresponding channel fading \cite{original-eurasip}--\cite{cluster}. As such, the mean STP only provides limited information on the performance of the cache-enabled networks since, as the mean of a random variable (i.e., STP), it cannot reflect the SIR variation or distribution among the individual links. In other words, the mean STP only answers the question of ``On average what fraction of users experience successful transmission (i.e., $\text{SIR}>\theta$)?''. In order to overcome this limitation and obtain a fine-grained general analysis on the network performance, we, inspired by \cite{meta-distribution}, adopt the meta distribution (MD) of the SIR as the performance metric, which is defined as the complementary cumulative distribution function (CCDF) of STP and answers the question ``What fraction of users can achieve the STP value of at least $y$ (an arbitrary percentage value)?'' \cite{meta-distribution}.

\subsection{Related Works}
The caching strategy in the cellular networks has been studied by utilizing stochastic geometry as the analyzing tool. In \cite{original-eurasip}, the analytical expressions for the average delivery rate and outage probability of the cache-enabled networks where the locations of BSs were modeled as Poisson point process (PPP) were investigated. In \cite{tier-level}, a closed-form expression of the cache hit probability for the cache-enabled heterogeneous cellular networks was derived. In addition, a sequential computation approach was proposed to obtain the optimal caching probability under the uniform signal-to-interference ratio (SIR) threshold, and an algorithm was proposed to achieve the sub-optimal solution under the non-uniform SIR threshold. The authors in \cite{ASE} proposed an optimal caching strategy to maximize the STP and the area spectral efficiency (ASE) of the cache-enabled networks. In addition, the relationship between the optimal caching probability and the network parameters was derived. In \cite{cluster}, a cluster-centric cellular network was proposed and a cooperative transmission strategy was designed to strike a balance between the content diversity and the transmission reliability.

The aforementioned works have provided the analytical results on the mean STP for the typical user without delving into the STP performance variation among the individual user-BS links for different BS distribution realizations. In order to overcome this limitation, a meta distribution based analysis framework needs to be developed for the cache-enabled networks.

As mentioned above, the foundation of meta distribution was laid in \cite{meta-distribution}, where the moments of the conditional success probability, the exact expression, bounds and approximation of the meta distribution for the cellular networks and bipolar networks were derived, respectively. The exact analytical expressions and the beta approximations of the meta distribution have since been obtained in various other scenarios, including the heterogeneous networks \cite{md-hetnet},  non-Poisson networks \cite{md-cellular-networks}, D2D communications \cite{md-D2D}, coordinated multipoint transmission \cite{md-bs-cooperation}, non-orthogonal multi-access \cite{md-NOMA} and fractional power control \cite{md-power-control}. Note that apart from the meta distribution, the mean local delay was also derived in \cite{md-hetnet}--\cite{md-power-control}.

It is well known that delay is an important performance metric reflecting service quality and network reliability \cite{rethinking}. In general, two kinds of delays exist in the wireless networks: the queuing delay, i.e., the waiting time for a packet in the service queues, and the transmission delay, i.e., the time taken by a packet for its successful transmission over the wireless link. The local delay, defined as the number of time slots taken until a successful transmission, is a key component of the transmission delay. The analysis of delay in a large wireless network requires the spatial and temporal analysis of the network \cite{massive-access}. In \cite{Haenggi-local-delay}, a mathematical framework was proposed for the derivation of the local delay by utilizing stochastic geometry. The work in \cite{local-delay} achieved the analytical expression of the local delay in the mobile Poisson networks. In \cite{delay-optimal}, the optimal power control policies were provided to minimize the local delay for different fading statistics. The authors in \cite{managing-interference-correlation} adopted two multi-access-control protocols, i.e., ALOHA and frequency-hop multiple access, to reduce the interference correlation, and corresponding parameters were optimized to obtain the minimization of local delay for both protocols. Note that the same set of interfering BSs may be seen by a user in different time slots (namely, the ``common randomness''), which introduces interference correlation and may increase the local delay. As such, the discontinuous transmission (DTX) strategy has been proposed to manage the inter-slot correlation by artificially introducing more randomness for the interference from the surrounding BSs. The local delay and the energy efficiency were analyzed for the wireless networks under PPP with random DTX scheme in \cite{weili}, and the results were extended to more general case of Poisson clustered process (PCP) in \cite{PCP}.

While the meta distribution and local delay in the traditional cellular networks have been investigated extensively, the meta distribution and the corresponding local delay in the cache-enabled networks still remain to be investigated. Compared with the traditional cellular networks, the meta distribution in the cache-enabled networks is also affected by the caching probability of the files and the main challenge is the cache capacity limitation of the BSs affecting the availability of the less popular files. In order to improve the network performance and enhance the file diversity, the random caching strategy can be adopted \cite{wen}. In general, the mean STP has also been employed for the performance evaluation of the cache-enabled networks under the random caching framework (e.g., \cite{cui} and \cite{wen}). In \cite{wen}, the authors analyzed the mean STP in the cache-enabled networks with random DTX and achieved the maximization of the mean STP in the static and high mobility scenarios. However, the local delay is not analyzed and the effect of the backhaul delay is not considered either. In this paper, therefore, the average system transmission delay \cite{Yaru}, incorporating the average delay caused by the transmission of a file to the users both from the BSs (for the BS cached files) and from the core network (for the non-BS cached files), is examined to characterize the effect of the local delay and the backhaul delay. Moreover, given the limitation of the mean STP as a metric, we in this paper go one step further by deriving the meta distribution of STP and the corresponding delay, and therefore provide a fine-grained general analysis framework for assessing the performance of cache-enabled networks.

\subsection{Contributions}
In the cache-enabled networks, maximization of the mean STP is obtained by optimizing the caching probability without considering the mean local delay \cite{wen}. However, a phase transition may occur where the mean local delay changes from the finite regime to the infinite regime \cite{md-D2D}. Therefore, the mean local delay is another critical performance metric to indicate the effectiveness of file transmission and as such is also adopted as an optimization objective in this paper.

We consider a downlink cache-enabled network where the random caching and random DTX schemes are applied to reduce the mean local delay. Our goals are (1) to maximize the STP under the BS cache capacity constraint and (2) to minimize the average system transmission delay by joint optimization of the random DTX and random caching parameters. The main contributions can be summarised as follows.
\begin{enumerate}
\item A fine-grain general analysis framework for STP and local delay of cache-enabled networks: We derive the $k$-th moment of the STP given a realization of the spatial locations of the BSs and the exact expression of the meta distribution of the SIR by utilizing stochastic geometry. The beta approximation of the meta distribution is also derived. Moreover, we derive the expression of the mean and variance of the local delay in the cache-enabled networks. From the numerical results, the critical values of the caching probability and the BS active probability are obtained and the relationship between the caching probability and the BS active probability is revealed.
\item Average STP and system transmission delay optimization schemes for cache-enabled networks: We consider the maximization of the mean STP and the minimization of the average system transmission delay, respectively. For the former, a convex optimization problem is formulated to maximize the mean STP under the BS cache capacity constraint. For the latter, a non-convex optimization problem is formulated to minimize the average system transmission delay by optimizing the BS active probability and the caching probability. By exploiting the optimality property, the problem is converted to a convex one and an iterative algorithm is proposed to achieve the optimal BS active probability and the caching probability.
\item Comprehensive simulation and analysis of the impact of key parameters: By numerical results, we demonstrate the effect of the key network parameters, i.e., the BS active probability, the caching probability and the SIR threshold, on the meta distribution and the mean local delay. In addition, the effect of the backhaul delay, the cache size and the Zipf exponent on the average system transmission delay under different caching strategies is also examined. In particular, it is confirmed that the network performance benefits from caching the most popular files when the backhaul delay is small while benefiting from a larger file diversity when the backhaul delay is large.
\end{enumerate}

\section{System Model}
We consider a downlink cache-enabled network. The BSs are assumed to follow a PPP $\Phi$ with density $\lambda$. The transmit power of BSs is fixed as $P$. The path loss function $\ell(x)=x^{-\alpha}$ is used, where $\alpha$ denotes the path loss exponent and $x$ is the distance between a user and its serving BS. The small-scale fading is assumed to be Rayleigh, i.e., $h\sim\mathcal{CN}(0,1)$. Without loss of generality, according to Slivnyak's theorem, we study the performance of the typical user $u_0$ located at the origin. We assume that the locations of BSs and users follow fixed, but arbitrary, PPP realizations.

Let $\mathcal{F}\triangleq \{1,2,\cdots,F\}$ denote a set of $F$ files in the network and all files are assumed to be of the same size\footnote{Note that the results can easily be extended to the case where the contents have different file sizes (e.g. by combining multiple files of different sizes to form files of equal size or splitting files of different sizes into segments of equal size) \cite{cui}.}. The file popularity distribution is assumed to be identical among all users. Let $p_f$ denote the probability that File $f$ is requested by a user, i.e., the popularity of File $f$ is $p_f$, and we have $\sum_{f=1}^{F}p_f=1$. In addition, we can always assume that $p_1\ge p_2 \ge \cdots \ge p_F$. Hence, the file popularity distribution can be expressed as $\mathbf{p}\triangleq \{p_1,p_2,\cdots,p_F\}$, which is assumed to be known as a priori. Note that the popularity of different files evolves at a relatively slow timescale and can be estimated in practice (e.g. by the machine learning \cite{machine-learning}). Each BS is equipped with a cache of size $C \le F$ to store $C$ different files from $\mathcal{F}$.

\subsection{Random Caching}
To improve the system performance and spatial diversity, the random caching strategy \cite{cui} is adopted. Let $q_{f}$ denote the probability that File $f$ is cached at a BS. Then, we have \cite{cui}
\begin{equation}\label{probability-constraint}
0\le q_{f} \le 1 \ \text{and}
\end{equation}
\begin{equation}\label{capacity-constraint}
\sum\limits_{f=1}^{F}q_{f}=C,
\end{equation}
which indicates that the sum of all the caching probabilities of the files cached at a BS is limited by the BS's storage capacity.

Consider the case that the typical user $u_0$ requests File $f$. If File $f$ is not stored in any BS, $u_0$ will download the corresponding file from the core network. Otherwise, $u_0$ is associated with the BS which provides the maximum biased received signal power among all BSs caching File $f$.

\subsection{Random DTX scheme}
We assume that the time is divided into equal slots and a transmission attempt requires a single time slot. In the random DTX scheme \cite{weili}, the DTX mode in each time slot is modeled as a Bernoulli trial with an parameter $\beta$ called BS active probability. That is, a BS in each time slot is temporally independently active with probability $\beta$ and is muted with probability $1-\beta$. Let $\Phi_{f}(t)$ and $\Phi_{-f}(t)$ denote the set of active BSs with/without caching File $f$ in time slot $t$, respectively. The interference at $u_0$ from the BSs with/without caching File $f$ at time slot $t$ is
\begin{equation}
I_{t,f}=\sum_{i\in\Phi_{f}\backslash x_0}\mathbf{1}(i\in\Phi_{f}(t))P\left|h_i\right|^2x_i^{-\alpha},
\end{equation}
\begin{equation}
I_{t,-f}=\sum_{i\in\Phi_{-f}}\mathbf{1}(i\in\Phi_{-f}(t))P\left|h_i\right|^2x_i^{-\alpha},
\end{equation}
where $\mathbf{1}(\cdot)$ denotes the indicator, $i$ is the index of BSs in $\Phi_f$ or $\Phi_{-f}$, and $x_0$ is the selected BS, $h_i$ and $x_i$ are the small-scale fading and the distance between the typical user $u_0$ and the $i$-th BS. We consider the interference-limited scenario and ignore the noise. Henceforth, the SIR at the typical user $u_0$ is
\begin{equation}
\text{SIR}=\frac{\mathbf{1}(0\in\Phi_{f}(t))P\left|h_0\right|^2r^{-\alpha}}{I_{t,f}+I_{t,-f}},
\end{equation}
where $h_0$ and $r$ denote the small-scale fading and the distance of the link between $u_0$ and its serving BS, respectively.

\subsection{Performance Metrics}
In this paper, we consider two performance metrics, i.e., the meta distribution and the mean local delay. Note that a retransmission occurs if a transmission in time slot $t$ fails. Therefore, the local delay is defined as the number of transmissions or retransmissions needed until a successful transmission occurs. Due to the application of the random DTX scheme, the transmission is considered to be successful if the serving BS is active and the SIR at $u_0$ exceeds a pre-defined threshold $\theta$. The corresponding probability is termed the successful transmission probability (STP, also called conditional STP, as it is conditioned on $\Phi$) given by
\begin{equation}
\mathcal{P}(\theta|\Phi)=\beta\mathbbm{P}\left[\text{SIR}>\theta\mid\Phi\right].
\end{equation}

The meta distribution of SIR \cite{meta-distribution} is defined as the CCDF of the conditional STP, which is given by
\begin{equation}\label{meta-distribution-equation}
\bar{F}_{\mathcal{P}}(y)\triangleq\mathbbm{P}\left[\mathcal{P}(\theta\mid\Phi)>y\right],\ y\in[0,1].
\end{equation}

Due to the ergodicity of the point processes, the meta distribution can physically be interpreted as the fraction of active links with the conditional STP greater than $y$ (with threshold $\theta$).

To obtain the analytical expression of the meta distribution, it is necessary to obtain the $k$-th moment of the conditional STP. When $u_0$ requests File $f$, we denote by $M_{k,f}$ the $k$-th moment of $\mathcal{P}$. The expression of meta distribution for File $f$ being requested can be derived by utilizing the Gil-Pelaez theorem \cite{Gil-Pelaez}. Since the exact form of the meta distribution is complicated, a simpler alternative method is adopted below where the beta distribution is utilized to approximate the meta distribution by matching the first and second moments.

We denote the mean local delay by $D$. Conditioned on $\Phi$, the local delay is geometrically distributed with parameter $\mathcal{P}(\theta|\Phi)$, we then have
\begin{equation}
\mathbbm{P}[D=d\mid\Phi]=(1-\mathcal{P}(\theta\mid\Phi))^{d-1}\mathcal{P}(\theta\mid\Phi),
\end{equation}
for $d=1,\cdots,\infty$. The mean of the geometrically distributed random variable $D$ conditioned on $\Phi$ is $\mathbbm{E}[D\mid\Phi]=\frac{1}{\mathcal{P}(\theta\mid\Phi)}$. The mean local delay can then be obtained by calculating the expectation with respect to $\Phi$ as follows
\begin{equation}\label{delay-definition}
\mathbbm{E}[D]=\mathbbm{E}_{\Phi}\left[\mathbbm{E}\left[D\mid\Phi\right]\right]=\mathbbm{E}_{\Phi}\left[\frac{1}{\mathcal{P}(\theta\mid\Phi)}\right].
\end{equation}

From (\ref{delay-definition}), it can be observed that the mean local delay is the -1-st moment of the conditional STP, i.e., $M_{-1,f}$. In this paper, our goal is therefore to obtain the maximization of the conditional STP or the minimization of the mean local delay by optimizing the caching probability and the BS active probability.

\section{Analytical Results}
In this section, we first analyze the meta distribution, then derive the mean local delay and obtain some useful insights.
\subsection{Meta Distribution}
In this subsection, we obtain the $k$-th moment of the conditional STP, i.e., $M_{k,f}$, when File $f$ is requested by $u_0$, then provide the exact expression of the meta distribution. In addition, the beta approximation of the meta distribution is derived.
\begin{Theorem}\label{theorem-moment}
When $u_0$ requests File $f$, the $k$-th moment of the conditional STP is given by
\begin{equation}\label{moment-equation}
\begin{split}
M_{k,f}=&q_f\left(q_f+\sum_{n=1}^{\infty}\binom{k}{n}(-1)^{n+1}\left(\delta(1-q_f)\beta^n\theta^{\delta}\text{B}(\delta,n-\delta)\right.\right.\\
&\left.\left.+\delta q_f\frac{(\beta\theta)^n}{n-\delta}F\left(n,n-\delta,n-\delta+1,-\theta\right)\right)\right)^{-1}.
\end{split}
\end{equation}
where $\delta=\frac{2}{\alpha}$, $F(a,b,c,z)=\frac{1}{\text{B}(b,c-b)}\int_{0}^{1}\frac{x^{b-1}(1-x)^{c-b-1}}{(1-zx)^a}\text{d}x$ and $\text{B}(a,b)=\int_{0}^{1}t^{a-1}(1-t)^{b-1}\text{d}t$.

\end{Theorem}

\emph{Proof:} See Appendix A.

Note that the mean STP and the mean local delay are the special cases of the $k$-th moment of the STP. The mean STP can be obtained when $k=1$ and the mean local delay can be obtained when $k=-1$. The meta distribution of the SIR defined in (\ref{meta-distribution-equation}) can then be obtained by applying the Gil-Pelaez theorem \cite{Zhong}, which is given by
\begin{equation}\label{md-equation}
\bar{F}_{\mathcal{P}_f}(y)=\frac{1}{2}+\frac{1}{\pi}\int_{0}^{\infty}\frac{\mathcal{J}\left(e^{-jt\log y}M_{jt,f}\right)}{t}\text{d}t,
\end{equation}
where
\begin{equation}\label{md-complex}
\begin{split}
M_{jt,f}&=q_f
\left(q_f+\sum_{n=1}^{\infty}\frac{\Gamma(jt+1)}{\Gamma(n+1)\Gamma(jt-n+1)}(-1)^{n+1}\left(\delta(1-q_f)\beta^n\theta^{\delta}\right.\right.\\
&\left.\left.\text{B}(\delta,n-\delta)+\delta q_f\frac{(\beta\theta)^n}{n-\delta}F\left(n,n-\delta,n-\delta+1,-\theta\right)\right)\right)^{-1},
\end{split}
\end{equation}
$\Gamma(x)=\int_{0}^{\infty}t^{x-1}e^{-t}\text{d}t$ and $\mathcal{J}(z)$ is the imaginary part of $z$. Since the numerical evaluation of (\ref{md-equation}) is cumbersome and it is difficult to obtain further insight, we now resort to a beta distribution to approximate the above meta distribution \cite{meta-distribution} by matching the first and second moments of STP, which can be easily obtained from the result in (\ref{moment-equation}) (or as special cases of (\ref{md-complex})):
\begin{equation}\label{mean-equation}
\begin{split}
M_{1,f}(\theta)=&q_f\left(q_f+\frac{q_f\beta\theta\delta}{1-\delta}F\left(1,1-\delta,2-\delta,-\theta\right)
+\delta(1-q_f)\beta\theta^{\delta}\text{B}(\delta,1-\delta)\right)^{-1} \text{and}
\end{split}
\end{equation}
\begin{equation}
\begin{split}
M_{2,f}(\theta)=&q_f\left(q_f+\frac{2q_f\beta\theta\delta}{1-\delta}F\left(1,1-\delta,2-\delta,-\theta\right)
-\frac{q_f(\beta\theta)^2\delta}{2-\delta}F\left(2,2-\delta,3-\delta,-\theta\right)\right.\\
&\left.+2(1-q_f)\delta\beta\theta^{\delta}\text{B}(\delta,1-\delta)-(1-q_f)\delta\beta^2\theta^{\delta}\text{B}(\delta,2-\delta)\right)^{-1}.
\end{split}
\end{equation}

By matching the variance and mean of the beta distribution, i.e., $M_{2,f}-M_{1,f}^2$ and $M_{1,f}$ (we have omitted $\theta$ here without causing confusion), the approximated meta distribution of the SIR is
\begin{equation}\label{ccu-md-approximation}
\bar{F}_{\mathcal{P}_f}(y)\approx 1-I_y\left(\frac{M_{1,f}\chi}{1-M_{1,f}},\chi\right),\ y\in[0,1],
\end{equation}
where $\chi=\frac{(M_{1,f}-M_{2,f})(1-M_{1,f})}{M_{2,f}-M_{1,f}^2}$ and $I_y(a,b)$ is the regularized incomplete beta function $I_y(a,b)\triangleq\frac{1}{\text{B}(a,b)}\int_{0}^{y}t^{a-1}(1-t)^{b-1}\text{d}t$.

\begin{figure}[htbp]
\begin{minipage}[t]{0.35\linewidth}
\centering
\includegraphics[width=3.5in]{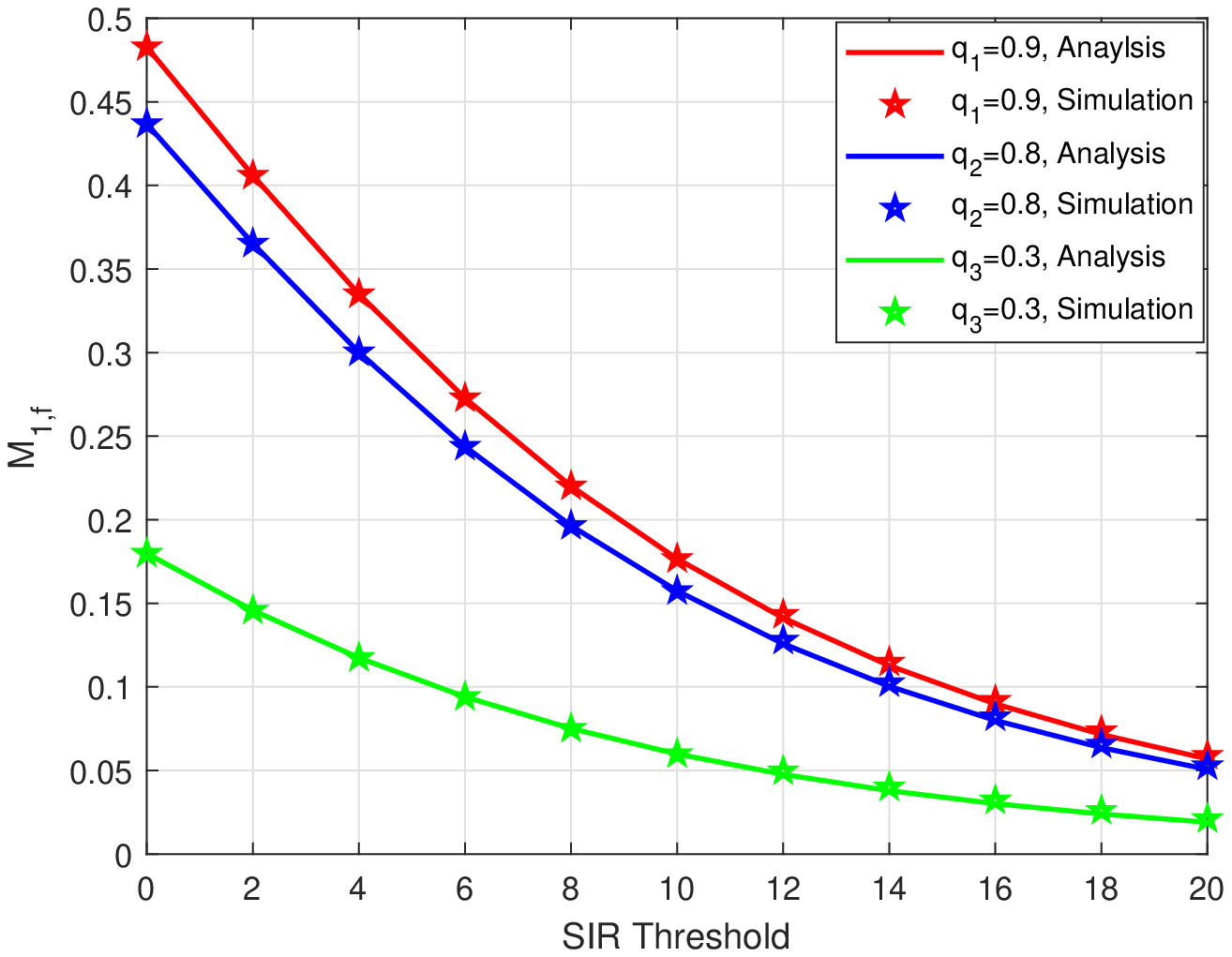}
\caption{$M_{1,f}$ (mean STP) as functions of SIR threshold $\theta$ with $F=10$ and $C=2$.}\label{moment-1-SIR-threshold}
\end{minipage}%
\hfill
\begin{minipage}[t]{0.5\linewidth}
\centering
\includegraphics[width=3.5in]{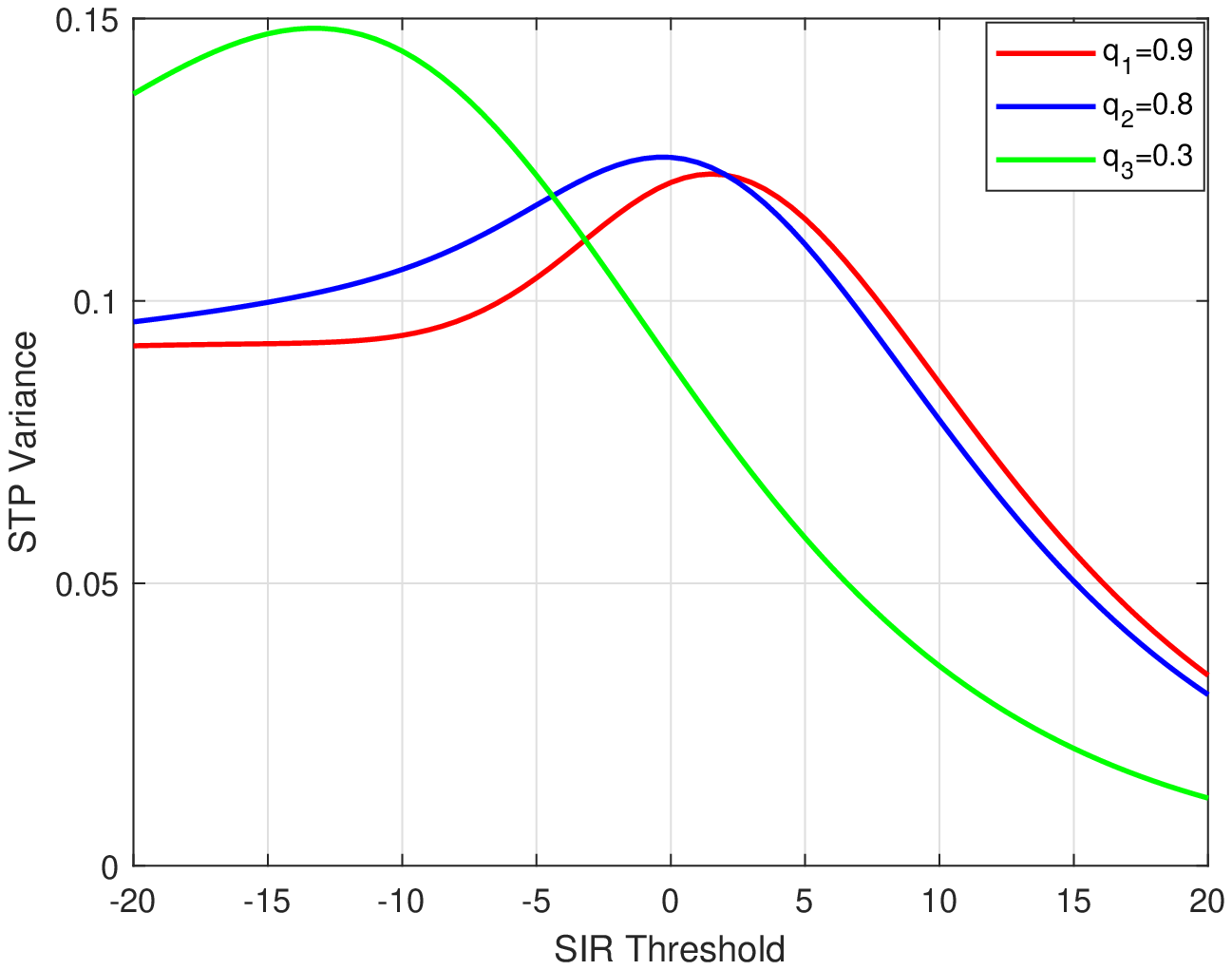}
\caption{STP variance as functions of SIR threshold $\theta$ with $F=10$ and $C=2$.}\label{variance-SIR-threshold}
\end{minipage}
\end{figure}

\begin{figure}[htbp]
\begin{minipage}[t]{0.35\linewidth}
\centering
\includegraphics[width=3.5in]{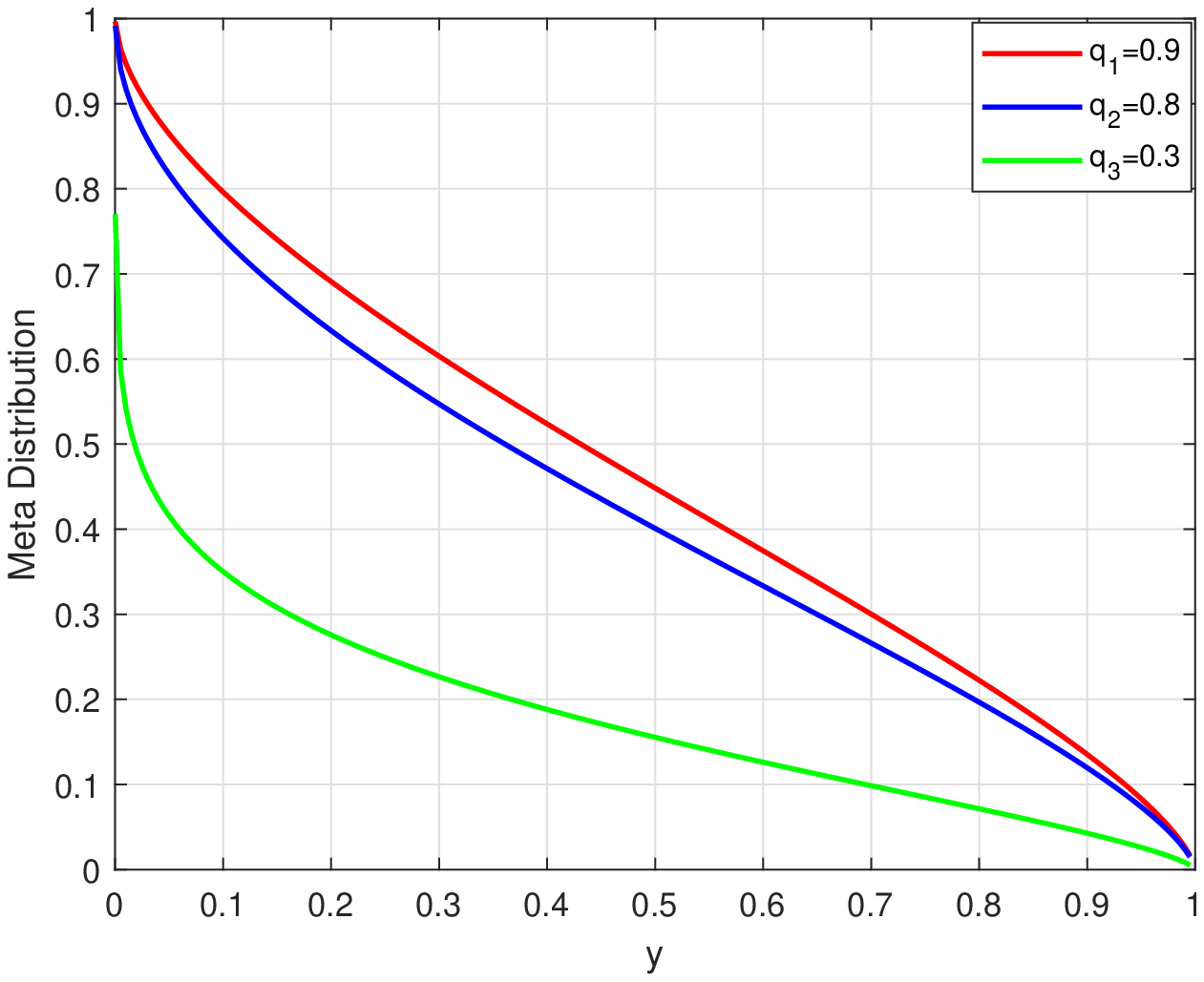}
\caption{Meta distribution with $F=10$ and $C=2$.}\label{meta-distribution-caching-probability}
\end{minipage}%
\hfill
\begin{minipage}[t]{0.5\linewidth}
\centering
\includegraphics[width=3.5in]{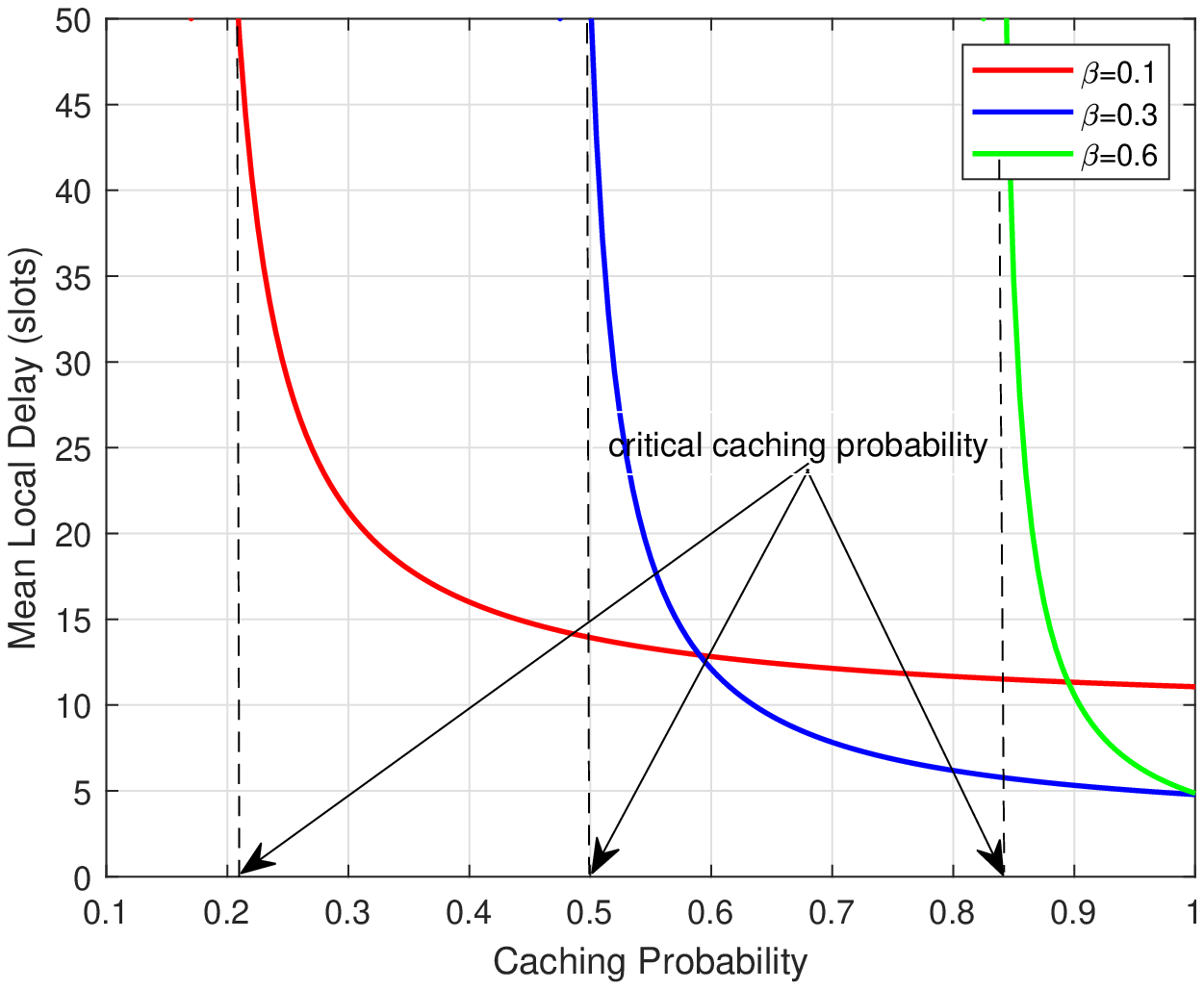}
\caption{Mean local delay as functions of caching probability $q_f$ under different active probabilities}\label{local-delay-caching-active}
\end{minipage}
\end{figure}

Fig. \ref{moment-1-SIR-threshold} and Fig. \ref{variance-SIR-threshold} plot the statistical information of the distribution of the STP, i.e., the mean ($M_{1,f}$) and the variance ($M_{2,f}-M_{1,f}^2$). From Fig. \ref{moment-1-SIR-threshold}, we can observe that the simulation results match the numerical results obtained from (\ref{mean-equation}) well, verifying the correctness of the theoretical analysis. From Fig. \ref{variance-SIR-threshold}, we observe that the variance of the STP (i.e., $M_{2,f}-M_{1,f}^2$) first increases with the SIR threshold $\theta$. After reaching its maximum, it starts to decrease. Note that the value of $\theta$ maximizing the variance increases with the caching probability, and the maximum variance decreases with the caching probability. Fig. \ref{meta-distribution-caching-probability} illustrates the meta distribution as a function of $y$ under different caching probabilities. It can be observed that the meta distribution decreases rapidly at first, then the variation of the meta distribution tends to become gentle. When $y$ approaches 1, the slope of the meta distribution becomes larger. In contrast, the mean STP is a constant when the network parameters are fixed. This phenomenon demonstrates the necessity of utilizing a more refined performance metric.

\subsection{Mean Local Delay}
In this subsection, we derive the closed-form expression of the mean local delay, then provide a simplified expression of the local delay under the special case where the SIR threshold is sufficiently large or small. While File $f$ is transmitted by the serving BS, the interfering BSs are divided into two groups: one group is formed by the active BSs with the requested file, and the other is formed by the active BSs without caching the requested file. By carefully handling the interference from these two groups, the expression of the mean local delay (as defined in (\ref{delay-definition})) can be obtained in the following theorem.
\begin{Theorem}\label{theorem-delay}
The mean local delay for File $f$ is
\begin{equation}\label{T-ld}
M_{-1,f}=\frac{q_f}{\beta(C_3q_f-C_1)},
\end{equation}
where
\begin{equation}
C_1=\delta(1-\beta)^{\delta-1}\beta\theta^{\delta}\text{B}(\delta,1-\delta),
\end{equation}
\begin{equation}
C_2=\frac{\delta\beta\theta}{1-\delta}F\left(1,1-\delta,2-\delta,-(1-\beta)\theta\right),
\end{equation}
\begin{equation}
C_3=1+C_1-C_2.
\end{equation}
\end{Theorem}
\emph{Proof:} See Appendix B.

From Theorem \ref{theorem-delay}, we can see that the mean local delay $M_{-1}$ is affected by the file-related parameters (i.e., the caching probability) and the network parameters (i.e., the active probability $\beta$ and the path loss exponent $\alpha$). The impacts of the file-related parameters and the network parameters are coupled. Moreover, the network jitter, i.e., the variance of the local delay, can be obtained. In order to obtain the network jitter, we first derive the -2-nd moment of the conditional STP for File $f$ as follows
\begin{equation}
\begin{split}
M_{-2,f}=&q_f\left(q_f-\sum_{n=1}^{\infty}(n+1)\left(\delta(1-q_f)\beta^n\theta^{\delta}\text{B}(\delta,n-\delta)\right.\right.\\
&\left.\left.+\delta q_f\frac{(\beta\theta)^n}{n-\delta}F\left(n,n-\delta,n-\delta+1,-\theta\right)\right)\right)^{-1}
\end{split}
\end{equation}

Then the network jitter can be given by
\begin{equation}\label{network-jitter-equation}
\mathrm{NJ}=\sum_{f=1}^{F}p_fM_{-2,f}-\left(\sum_{f=1}^{F}p_fM_{-1,f}\right)^2.
\end{equation}

The above expressions can be simplified for the low and high data rate regimes (i.e., when $\theta\rightarrow 0$ and $\infty$), offering more insight. This, however, has been left out here due to the page limit.

Note that the mean local delay is independent of the BS density $\lambda$ and the transmit power $P$. The reason is as follows. When the density of BSs in the network increases, the distance between $u_0$ and its serving BS becomes smaller, resulting in a stronger signal received by $u_0$. Meanwhile, the interference received by $u_0$ becomes stronger too. Overall, the statistics of STP remains unchanged.

In addition, it can be proved that the mean local delay is an increasing function of the SIR threshold $\theta$. The reason can be explained as follows. When the SIR threshold $\theta$ increases, the probability that File $f$ is successfully transmitted by the BS in a certain time slot decreases. Therefore, a larger number of retransmissions are needed until the transmission succeeds. Since the local delay reaches infinity when the SIR threshold $\theta$ exceeds a certain value, we observe that a phase transition occurs when the mean local delay changes from finite to infinity \cite{md-D2D}. The corresponding value of $\theta$ is called the critical value, denoted by $\theta_c$. When the caching probability $q_f$ and the active probability $\beta$ are fixed, the critical value of the SIR threshold can be obtained by letting the denominator of (\ref{T-ld}) be equal to 0:
\begin{equation}
\begin{split}
&q_f-\delta(1-q_f)(1-\beta)^{\delta-1}\beta\theta_c^{\delta}\text{B}(\delta,1-\delta)
-\frac{q_f\delta\beta\theta_c}{1-\delta}F\left(1,1-\delta,2-\delta,-(1-\beta)\theta_c\right)=0.
\end{split}
\end{equation}

The above equation indicates that the local delay for File $f$ $M_{-1,f}\rightarrow\infty$ once $\theta>\theta_c$, given a certain caching probability $q_f$ and active probability $\beta$.

Next, we investigate the effect of the caching probability and the active probability on the mean local delay $M_{-1}$. Fig. \ref{local-delay-caching-active} plots the mean local delay as a function of active probability under different caching probabilities. From Fig. \ref{local-delay-caching-active}, it can be observed that the mean local delay decreases with the caching probability. The minimum value of local delay can be achieved when the caching probability is 1. This can be explained as that when the caching probability becomes larger, the distance between $u_0$ and its serving BS is smaller, leading to stronger signal received by $u_0$. Hence, the corresponding SIR increases and the mean local delay decreases. It can also be observed that the mean local delay approaches infinity when the caching probability is below a certain value. This is because that the distance between $u_0$ and its serving BS is so large that the file cannot be transmitted and decoded successfully. Under this situation, the file is retrieved from the core network through the backhaul link. Such a ``critical caching probability'' $q_{c}$ can be obtained by letting the denominator of (\ref{T-ld}) be equal to 0. Therefore, the critical caching probability for any file can be determined as
\begin{equation}\label{critical-caching-probability-equation}
q_{c}\sim\frac{C_1}{C_3}.
\end{equation}

From Fig. \ref{local-delay-caching-active}, it is also noteworthy that the critical caching probability $q_{c}$ increases with the BS active probability. This phenomenon indicates that the caching probability needs to be large in order to make the local delay finite in the large active probability regime.

\begin{figure}[htbp]
\begin{minipage}[t]{0.35\linewidth}
\centering
\includegraphics[width=3.5in]{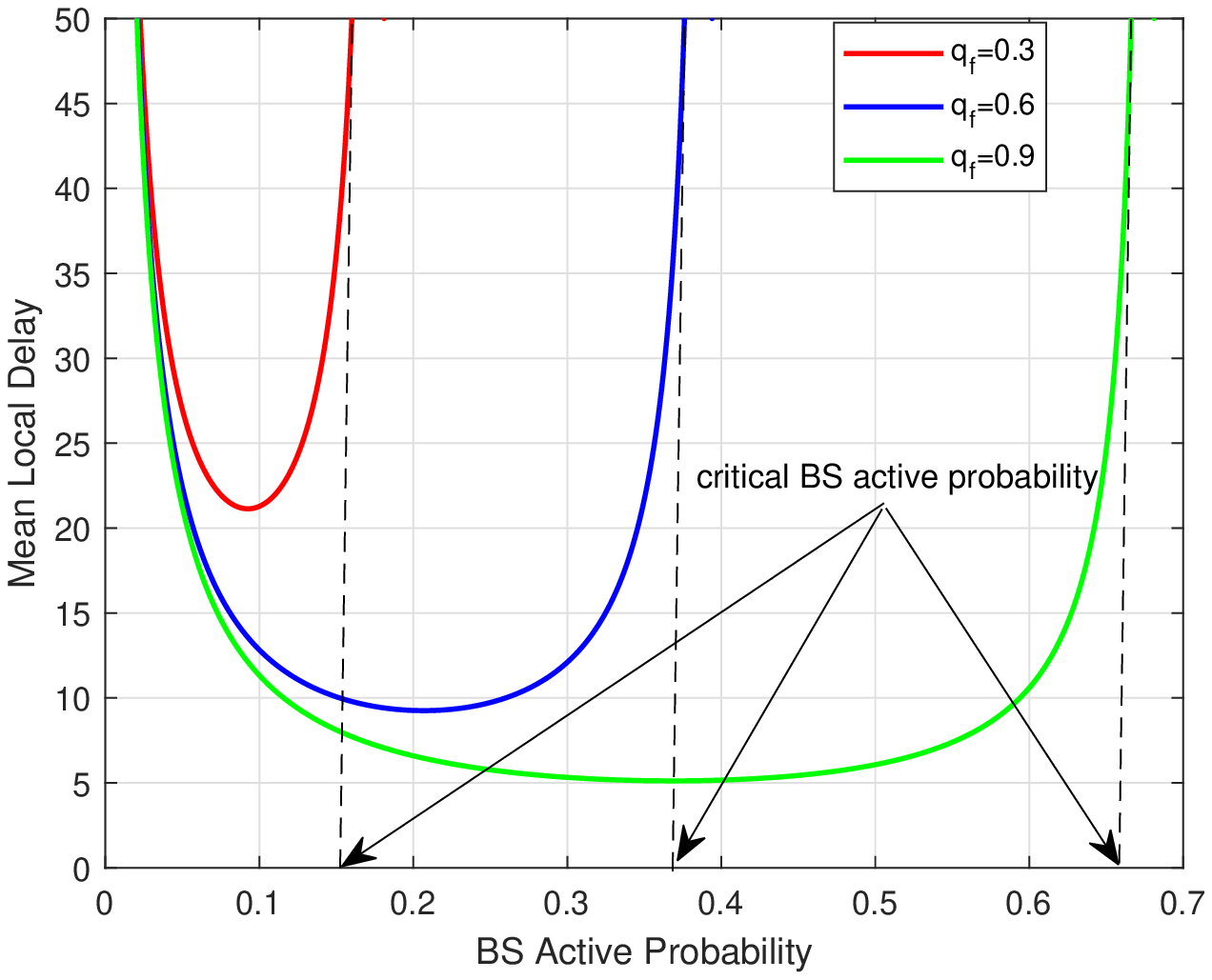}
\caption{Mean local delay as functions of active probability $\beta$ under different caching probabilities}\label{local-delay-active-caching}
\end{minipage}%
\hfill
\begin{minipage}[t]{0.5\linewidth}
\centering
\includegraphics[width=3.5in]{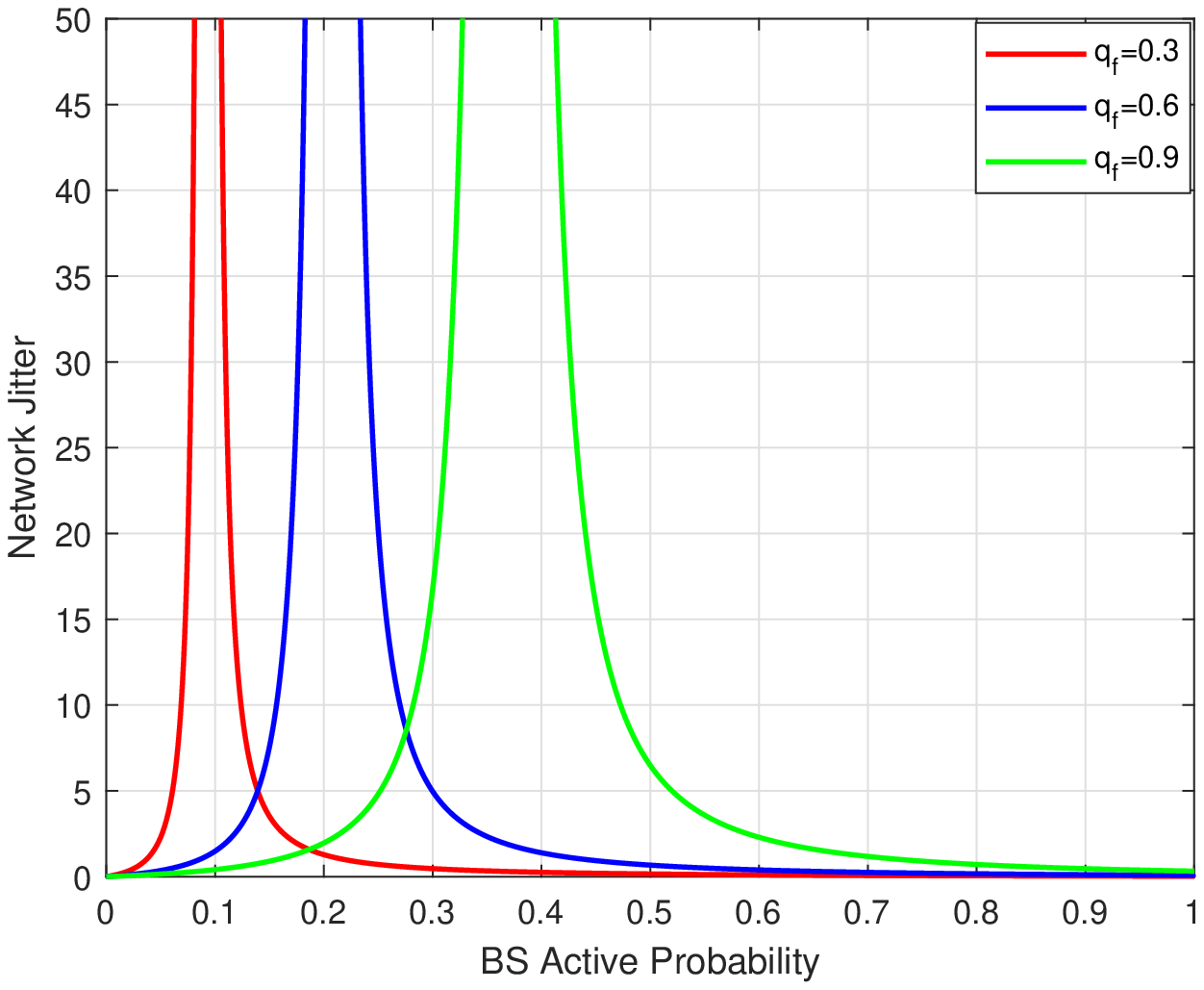}
\caption{Network jitter as functions of active probability $\beta$ under different caching probabilities}\label{network-jitter-active-caching}
\end{minipage}
\end{figure}

Fig. \ref{local-delay-active-caching} illustrates the effect of the active probability $\beta$ on the mean local delay under different caching probabilities. From Fig. \ref{local-delay-active-caching}, we observe that the mean local delay approaches infinity when the active probability $\beta$ decreases to zero or increases to a certain value. This value is defined as the ``critical active probability'' $\beta_c$ and it can be obtained following a similar method to the derivation of $q_{c}$. Note, however, that there exists an optimal value of active probability $\beta^{*}$ within $(0,\beta_c)$. The mean local delay increases with the gap $\left|\beta-\beta^*\right|$. The reason can be explained as follows. When $\beta<\beta^*$, the active probability for the serving BS becomes smaller, leading to the increasing number of retransmission needed until the transmission succeeds. When $\beta>\beta^*$, the interference received by $u_0$ becomes larger, leading to the decrease of SIR. In addition, $u_0$ in this case (i.e., $\beta>\beta^*$) is more likely to be interfered by a similar set of BSs in different time slots. The interference temporal correlation together with the increasing interference will reduce the probability that the file is transmitted by the BS successfully.
Fig. \ref{network-jitter-active-caching} illustrates the effect of the active probability $\beta$ on the network jitter $\mathrm{NJ}$ defined in (\ref{network-jitter-equation}) under different caching probabilities. The fluctuation of the local delay can be reflected by the network jitter. A larger network jitter corresponds to a larger fluctuation of the local delay and vice versa \cite{deng-SINR}. It can be observed that the network jitter decreases to zero when $\beta$ approaches zero or one. The reason can be explained as follows. With random DTX, each BS is active with probability $\beta$, leading to the fluctuation of the signal and interference power received by $u_0$. Therefore, the STP variance increases and so does the fluctuation in the number of retransmissions needed until the transmission succeeds. For example, the lowest mean local delay in Fig. 5 corresponds to the most random SIR or STP, leading to the highest variance of local delay, as reflected by the highest network jitter ($\mathrm{NJ}$) in Fig. \ref{network-jitter-active-caching}.

\section{Optimization of STP and Average System Transmission Delay}

\subsection{Optimization of Mean STP}
In this subsection, we focus on the maximization of the STP by optimizing the caching probability and the BS active probability. Different from \cite{wen} which analyzed the probability that a file is successfully transmitted before a predefined deadline, we focus on the mean STP of a file in a certain time slot. Furthermore, the optimization of the mean STP is from the perspective of meta distribution, which offers a much more general analysis framework.  The optimization problem can be formulated as follows.

\emph{Problem 1 (Optimization of mean STP):}
\begin{equation}
\max\limits_{\mathbf{q},\beta}\ M_{1}(\mathbf{q},\beta)\triangleq\sum_{f=1}^{F}p_fM_{1,f}(\mathbf{q},\beta)
\end{equation}

\centerline{$\text{s.t.}\ \ (\ref{probability-constraint}), (\ref{capacity-constraint}).$}

Problem 1 is the maximization of a non-convex function within a convex set. In general, to obtain a globally optimal solution for a non-concave problem is difficult. Here we achieve the globally optimal solution for the problem by exploring the optimal structures of $M_{1}(\mathbf{q},\beta)$. It can be easily verified that $ M_{1}(\mathbf{q},\beta)$ is a monotonically increasing function of the BS active probability $\beta$. Given a fixed $\beta$, the problem turns out to be concave and the optimal caching probability can be obtained by utilizing the Karush-Kuhn-Tucker (KKT) conditions. The Lagrange function of Problem 1 can be written as follows
\begin{equation}
\begin{split}
L(\mathbf{q},\tau)=M_{1}(\mathbf{q},\beta)+\tau\left(C-\sum_{f=1}^{F}q_f\right),
\end{split}
\end{equation}

By letting the derivative of the above Lagrange function equal to 0, the optimal caching probability can be obtained as follows
\begin{equation}
q_f^*=\min\left\{\max\left\{\frac{1}{\Psi_1}\sqrt{\frac{p_f\Psi_2}{\tau}}-\frac{\Psi_2}{\Psi_1},0\right\},1\right\}
\end{equation}
where
\begin{equation}
\Psi_1=1-\delta\beta\theta^{\delta}\text{B}\left(\delta,1-\delta\right)+\frac{\delta\beta\theta}{1-\delta}
F\left(1,1-\delta,2-\delta,-\theta\right),
\end{equation}
\begin{equation}
\Psi_2=\delta\beta\theta^{\delta}\text{B}\left(\delta,1-\delta\right).
\end{equation}

and $\tau$ satisfies
\begin{equation}
\sum_{f=1}^{F}\min\left\{\max\left\{\frac{1}{\Psi_1}\sqrt{\frac{p_f\Psi_2}{\tau}}-\frac{\Psi_2}{\Psi_1},0\right\},1\right\}=C.
\end{equation}
\subsection{Optimization of Average System Transmission Delay}
In this subsection, the average system transmission delay is analyzed. The average system transmission delays are different for the files cached in the BSs and those not cached (e.g., \cite{Yaru}). When $u_0$ requests a file which happens to be cached in the BS (i.e., $q_f>0$), $u_0$ can directly obtain the corresponding file from the BS in which case only the local delay, denoted by $M_{-1,f}$, is involved. However, when the requested file is not cached in any BS (i.e., $q_f=0$), $u_0$ needs to retrieve the corresponding file from the core network through the backhaul. Hence, the average system transmission delay now includes both the mean local delay and the backhaul delay, denoted by $D_{nc}$. Since the DTX scheme is employed, only the interference from the active BSs will be received by $u_0$. The average system transmission delay $D_{ai}(\mathbf{q},\beta)$ can therefore be expressed as
\begin{equation}\label{D-ai-equation}
D_{ai}(\mathbf{q},\beta)=\sum_{f=1}^{F}p_f\left(\mathbf{1}(q_f>0)M_{-1,f}+\mathbf{1}(q_f=0)D_{nc}\right),
\end{equation}
where $\mathbf{1}(\cdot)$ denotes the indictor function, the mean local delay for the files cached in the BSs is given by (\ref{T-ld}), and the delay for the files not cached in any BS consists of the backhaul delay and the local delay between $u_0$ and its nearest BS, which is given by
\begin{equation}
D_{nc}=\frac{1}{\beta(C_3-C_1)}+\xi
\end{equation}
where $\xi$ is the backhaul delay.

Next, we we obtain the minimization of the average system transmission delay by optimizing the caching probability $\mathbf{q}$ and the active probability $\beta$. The optimization problem is formulated as follows.

\emph{Problem 2 (Optimization of average system transmission delay):}
\begin{equation}\label{problem-2-equation}
\min\limits_{\mathbf{q},\beta}\ D_{ai}(\mathbf{q},\beta)
\end{equation}
$\centerline{\text{s.t.}\ \ (\ref{probability-constraint}), (\ref{capacity-constraint}),}\\$
where, as before, (\ref{probability-constraint}) is the probability constraint and (\ref{capacity-constraint}) is the capacity constraint. In order to facilitate the optimization process, we explore the optimality property of this problem first:
\begin{enumerate}
\item The network parameters affect the average system transmission delay of all files while the caching probability only affects the average system transmission delay of the corresponding file.
\item Given a fixed active probability $\beta$, there exists $F_c^*\in\left[C,\min\left(\left\lceil\frac{C}{q_c}\right\rceil-1,F\right)\right]$ such that the the caching probabilities of the files stored in the BSs exceed the critical caching probability, i.e., $q_f>q_c, f\in\left[1,F_c^*\right]$ and $q_f=0, f\in\left(F_c^*,F\right]$, where $q_c$ is determined by (\ref{critical-caching-probability-equation}). Given a fixed $\mathbf{q}$, there exists an optimal $\beta^*\in\left(0,\beta_c\right)$ such that the minimum local delay is achieved.
\item When $F_c^*=C$, we have $q_f=1, f\in\left[1,C\right]$ and $q_f=0, f\in\left(C,F\right]$, indicating that the optimal caching strategy reduces to the ``most popular content (MPC)'' scheme \cite{original-eurasip}.
\end{enumerate}

Due to the existence of the indicator function, it is difficult to obtain the derivative of the objective function (\ref{D-ai-equation}). The gradient projection method \cite{cui} or interior point method \cite{boyd} cannot be directly applied. Therefore, an equivalent problem is constructed by utilizing the optimality property of Problem 2. Since $F_c^*$ files are cached in the BSs, an auxiliary variable $F_c$ is introduced and the objective function of Problem 2 can be rewritten as
\begin{equation}
D_{ai}(\mathbf{q},\beta)=D_{1}(\mathbf{q},F_c)+D_{2}(F_c),
\end{equation}
where $D_{1}(\mathbf{q},F_c)=\sum_{f=0}^{F_c}p_fM_{-1,f}$ and $D_{2}(F_c)=\sum_{f=F_c+1}^{F}p_fD_{nc}$, respectively. We can see that $D(\mathbf{q},\beta)$ is differentiable with respect to $\mathbf{q}$ and $\beta$. The equivalent problem can then be formulated as follows.

\emph{Problem 3 (The Equivalent Problem of Problem 2):}
\begin{align}
\min\limits_{\mathbf{q},\beta,F_c}\ &D_{ai}(\mathbf{q},\beta,F_c)\label{Problem 2}\\
\text{s.t.}\ \ C<F_c&<\min\left\{\left\lceil\frac{C}{q_c}\right\rceil-1,F\right\}\tag{\ref{Problem 2}{a}}\label{Problem 2-1}\\
\beta_c&<\beta<1\tag{\ref{Problem 2}{b}}\label{Problem 2-2}\\
q_c&<q_f<1\tag{\ref{Problem 2}{c}}\label{Problem 2-3}\\
\sum_{f=1}^{F_c}&q_f=C,\tag{\ref{Problem 2}{d}}\label{Problem 2-4}
\end{align}
where constraint (\ref{Problem 2-1}) limits the search space of $F_c$. The probability constraint (\ref{Problem 2-3}) and capacity constraint (\ref{Problem 2-4}) are refined from (\ref{probability-constraint}) and (\ref{capacity-constraint}) by utilizing the optimality property. From (\ref{Problem 2}), we can observe that two types of variables are to be determined. One is the continuous variables $\mathbf{q}$ and $\beta$ and the other is the discrete variables $F_c$. $\beta$ and $\mathbf{q}$ can be updated iteratively until a stationary point is reached. Specifically, two sub-problems are formulated to obtain the optimal $\beta$ and $\mathbf{q}$, respectively.

\emph{Sub-Problem 1 (Optimization of Caching Probability for Given Active Probability):}
\begin{equation}
D_{ai}^*(\mathbf{q},\beta,F_c)\triangleq\min\limits_{F_c}\ D_1^*(F_c)+D_2(F_c)\\
\end{equation}
$\centerline{\text{s.t.}\ \ (\ref{Problem 2-1})}\\$
where
\begin{equation}
D_1^*(F_c)\triangleq\min\limits_{\mathbf{q}}D_1(\mathbf{q},F_c)\label{subproblem-1}\\
\end{equation}
$\centerline{\text{s.t.}\ \ (\ref{Problem 2-3}),\ (\ref{Problem 2-4}).}$

Note that $F_c$ can be determined by exhaustive search with complexity $\mathcal{O}(F)$. For a given $F_c$, it can be easily verified that the second-order derivative of $D(\mathbf{q},\beta,F_c)$ with respect to $\mathbf{q}$ is positive within the feasible region. Therefore, the problem is convex and satisfies Slater's condition, indicating that a strong duality holds. Then we can obtain the optimal caching probability in the following lemma.
\begin{Lemma}\label{lemma-laplace}
The optimal caching probability can be obtained by utilizing the KKT conditions as follows
\begin{equation}
q_f^*(t)=\min\left\{\max\left\{\frac{1}{C_3}\sqrt{\frac{-p_fC_1}{\eta\beta}}+\frac{C_1}{C_3},q_c\right\},1\right\},
\end{equation}
where $\eta$ satisfies
\begin{equation}
\sum_{f=1}^{F}\min\left\{\max\left\{\frac{1}{C_3}\sqrt{\frac{-p_fC_1}{\eta\beta}}+\frac{C_1}{C_3},q_c\right\},1\right\}=C.
\end{equation}
\end{Lemma}

\emph{Proof:} See Appendix D.

\emph{Sub-Problem 2 (Optimization of BS Active Probability for Given Caching Probability):}
\begin{equation}\label{subproblem-2}
\min\limits_{\beta}\ D_{ai}(\mathbf{q},\beta,F_c)
\end{equation}
$\centerline{\text{s.t.}\ \ (\ref{Problem 2-2}).}$

It is difficult to determine the convexity of the problem. Since the derivative of the objective function can be achieved, we can employ the gradient projection method \cite{cui} to obtain the locally optimal BS active probability. The derivative of the objective function is given by
\begin{equation}
\begin{split}
&\frac{\partial D_{ai}(\mathbf{q},\beta,F_c)}{\partial\beta}=q_f\left(2\Omega_2\beta-\Omega_1\sum_{n=0}^{\delta-1}\binom{\delta-1}{n}(n+2)\right.\\
&\left.(-\beta)^{n+1}-q_f\right)\left(\Omega_1(1-\beta)^{\delta-1}\beta-\Omega_2\beta+q_f\right)^{-2},
\end{split}
\end{equation}
where
\begin{equation}
\Omega_1=q_f\delta(1-\beta)^{\delta-1}\beta\theta^{\delta}\text{B}(\delta,1-\delta),
\end{equation}
\begin{equation}
\Omega_1=\frac{q_f\delta\theta}{1-\delta}F(1,1-\delta,2-\delta,-(1-\beta)\theta).
\end{equation}

Sub-problems 1 and 2 can then be carried out alternately in an iterative manner. In each iteration, given $\beta$, the critical caching probability $q_c$ is first obtained and the search range of the number of cached files $F_c$ can be determined, within which the exhaustive search is performed over $F_c$ and the KKT conditions are utilized to obtain the optimal caching probability for each $F_c$. Note that a bisection search is required to find $\eta$. Letting the upper and lower bounds of the search space for $\eta$ be $\eta_u$ and $\eta_l$, respectively, and assuming that the search precision is set to be $\eta_{a}$, the complexity can then be characterized by $\mathcal{O}\left(\left(\min\left\{\left\lceil\frac{C}{q_c}\right\rceil-1,F\right\}-C\right)\log\frac{\eta_u-\eta_l}{\eta_{a}}\right)$. Given caching probability $\mathbf{q}$, the critical BS active probability $\beta_c$ is first determined and the gradient projection method is applied to obtain the optimal BS caching probability $\beta$. The complete process is summarised below as Algorithm 1.

\begin{algorithm}
\begin{footnotesize}
\caption{Solution of Problem 2: Equation (\ref{problem-2-equation})} 
\hspace*{0.02in} {\bf Input:} 
Number of Files $F$, Cache size $C$.\\
\hspace*{0.02in} {\bf Output:} 
Optimal caching probability $\mathbf{q}^{*}$.\\
\hspace*{0.02in} {\bf Initialize:} 
set $\chi=0$.
\begin{algorithmic}[1]
\State $\mathbf{repeat}$
\State Obtain the critical caching probability $q_c$
\For{$F_c=C$ to $\min\left\{\left\lceil\frac{C}{q_c}\right\rceil-1,F\right\}$}
\State Obtain $F_c^*$ and $\mathbf{q}^*$ by solving the optimization in (\ref{subproblem-1}) using the KKT conditions.
\If {$D_{ai}^*(\mathbf{q},\beta,F_c)<D_1^*(F_c)+D_2(F_c)$}
\State $D_{ai}^*(\mathbf{q},\beta,F_c)\leftarrow D_1^*(F_c)+D_2(F_c)$ and $\mathbf{q}^*\leftarrow\mathbf{q}^*(F_c^*)$
\EndIf
\State \textbf{end if}
\EndFor
\State $\textbf{end for}$
\State Obtain the critical BS active probability $\beta_c$
\State Obtain $\beta^*$ by solving the optimization in (\ref{subproblem-2}) using the gradient projection method.
\State $\chi\rightarrow\chi+1$
\State $\mathbf{until}$ convergence criterion is satisfied.
\State \Return the optimal caching probability $\mathbf{q}^{*}$.
\end{algorithmic}
\end{footnotesize}
\end{algorithm}
\section{Simulation Results}
In this section, we first illustrate the effect of the critical network parameters, i.e., the active probability and the SIR threshold, on the meta distribution and mean local delay. Unless otherwise stated, the BS transmit power is $P=23$dBm, the BS density is $\lambda=10^{-4}/\text{m}^2$, the user density is $\lambda_u=3\times10^{-4}/\text{m}^2$, and the path loss exponent $\alpha=3$. We assume the popularity $p_f$ of the files satisfies the Zipf distribution, i.e., $p_{f}=\frac{f^{-\gamma}}{\sum_{f=1}^{F}f^{-\gamma}}$, where $\gamma$ is the Zipf exponent which reflects the skewedness of the file popularity distribution. Also, the files are ranked according to their popularity: $p_1>p_2 ... >p_{30}$.

\begin{figure}[htbp]
\begin{minipage}[t]{0.35\linewidth}
\centering
\includegraphics[width=3.5in]{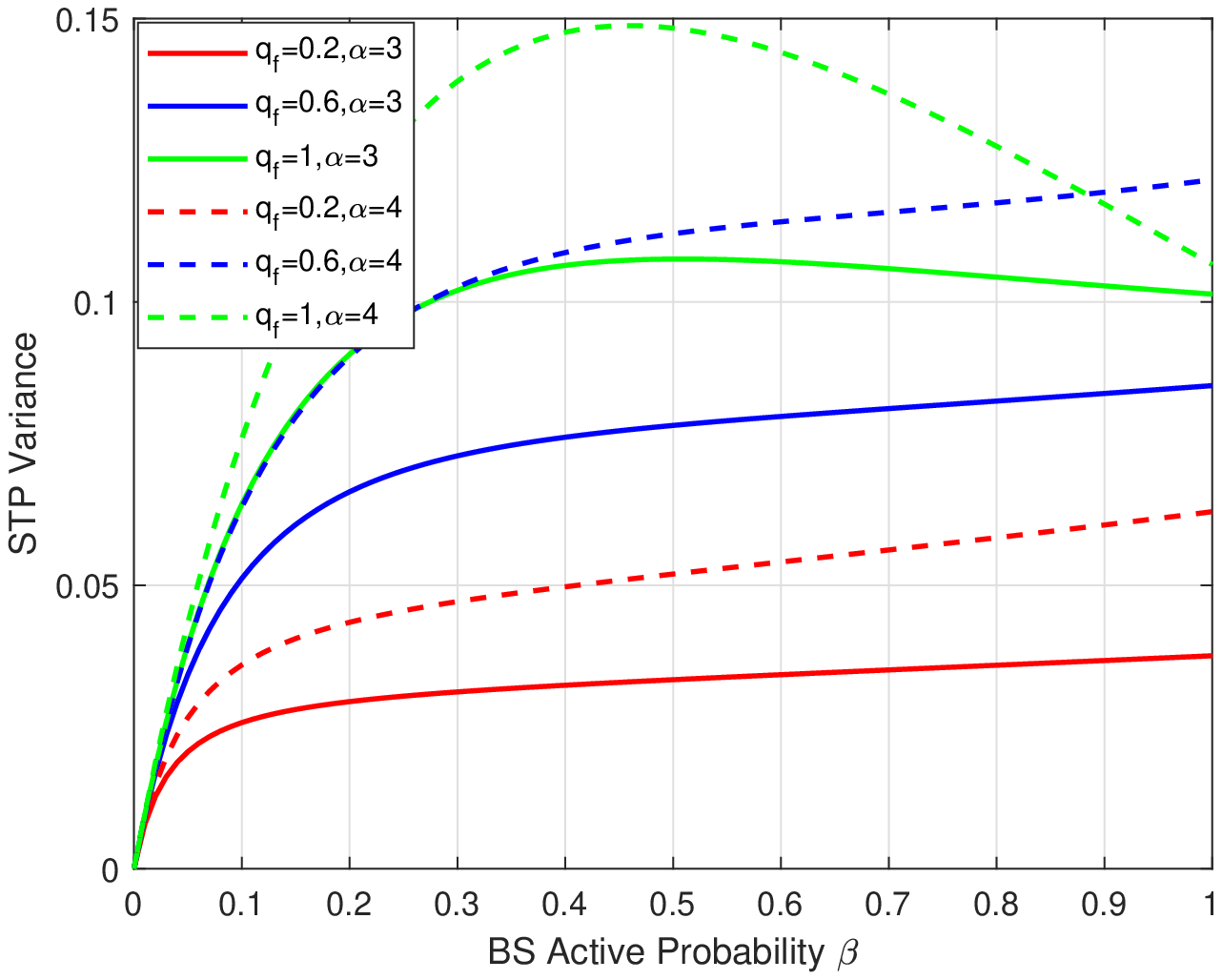}
\caption{STP variance as functions of BS active probability $\beta$.}\label{variance-active-probability}
\end{minipage}%
\hfill
\begin{minipage}[t]{0.5\linewidth}
\centering
\includegraphics[width=3.5in]{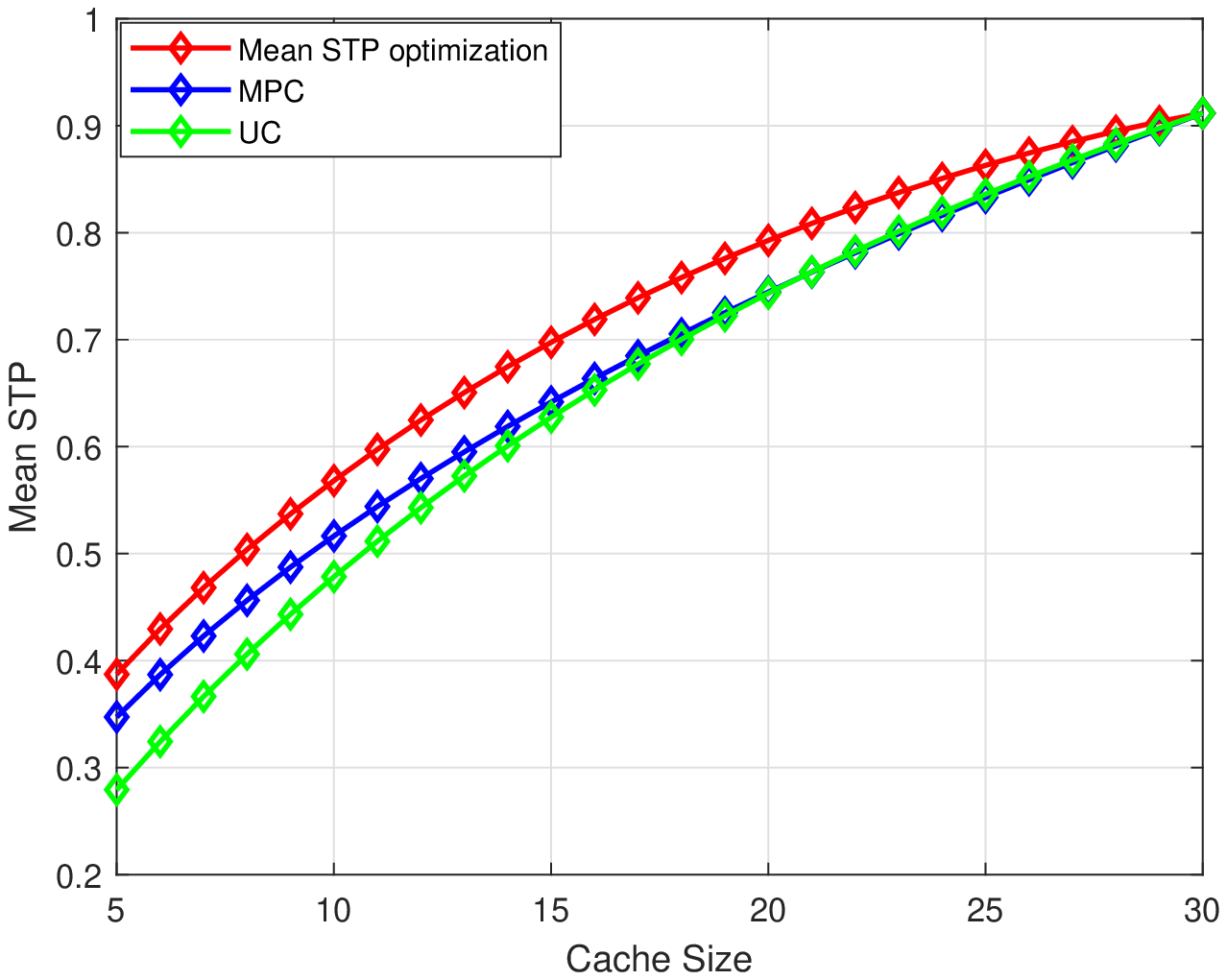}
\caption{Mean STP as functions of cache size $C$ under different caching strategies.}\label{STP-cache-size}
\end{minipage}
\end{figure}

Fig. \ref{variance-active-probability} plots the variance of the STP as functions of the BS active probability. The performance fluctuation can be reflected by the STP variance. A large variance corresponds to a large performance fluctuation and vice versa \cite{deng-SINR}. From Fig. \ref{variance-active-probability}, it can be observed that there exists a maximum variance when the caching probability $q$ is relatively large, i.e., $q_f=$1. Given $q_f=$0.2 or 0.6, the STP variance increases with $\beta$ rapidly at start, then the variation tends to be gentle. The reason can be explained as follows. With random DTX, each BS is activated with probability $\beta$, leading to the fluctuation of the signal and interference power received by $u_0$. Moreover, when $q_f=$0.2 or 0.6, the STP variance increases with the path loss exponent $\alpha$, indicating that the performance fluctuation increases when the inter-cell interference is reduced and each BS is more likely to be isolated.

Fig. \ref{STP-cache-size} presents the impact of cache size on the mean STP under different caching strategies, i.e., the proposed method, the MPC and uniform caching (UC) \cite{UC} strategies. For the MPC strategy, only the most popular files are cached in the BSs, i.e., $q_f=$1 for $f\in[1,C]$ and $q_f=$0 for $f\in[C+1,F]$. For the UC strategy, all files are cached in the BSs with equal probabilities, i.e., $q_f=C/F$ for $f\in[1,F]$. It can be observed that the proposed optimization method always outperforms the other two strategies. In addition, the mean STP for the MPC strategy outperforms the UC strategy when the cache size is small and the performance gap between the MPC and UC strategies becomes smaller when the cache size increases.

\begin{figure}[htbp]
\begin{minipage}[t]{0.35\linewidth}
\centering
\includegraphics[width=3.5in]{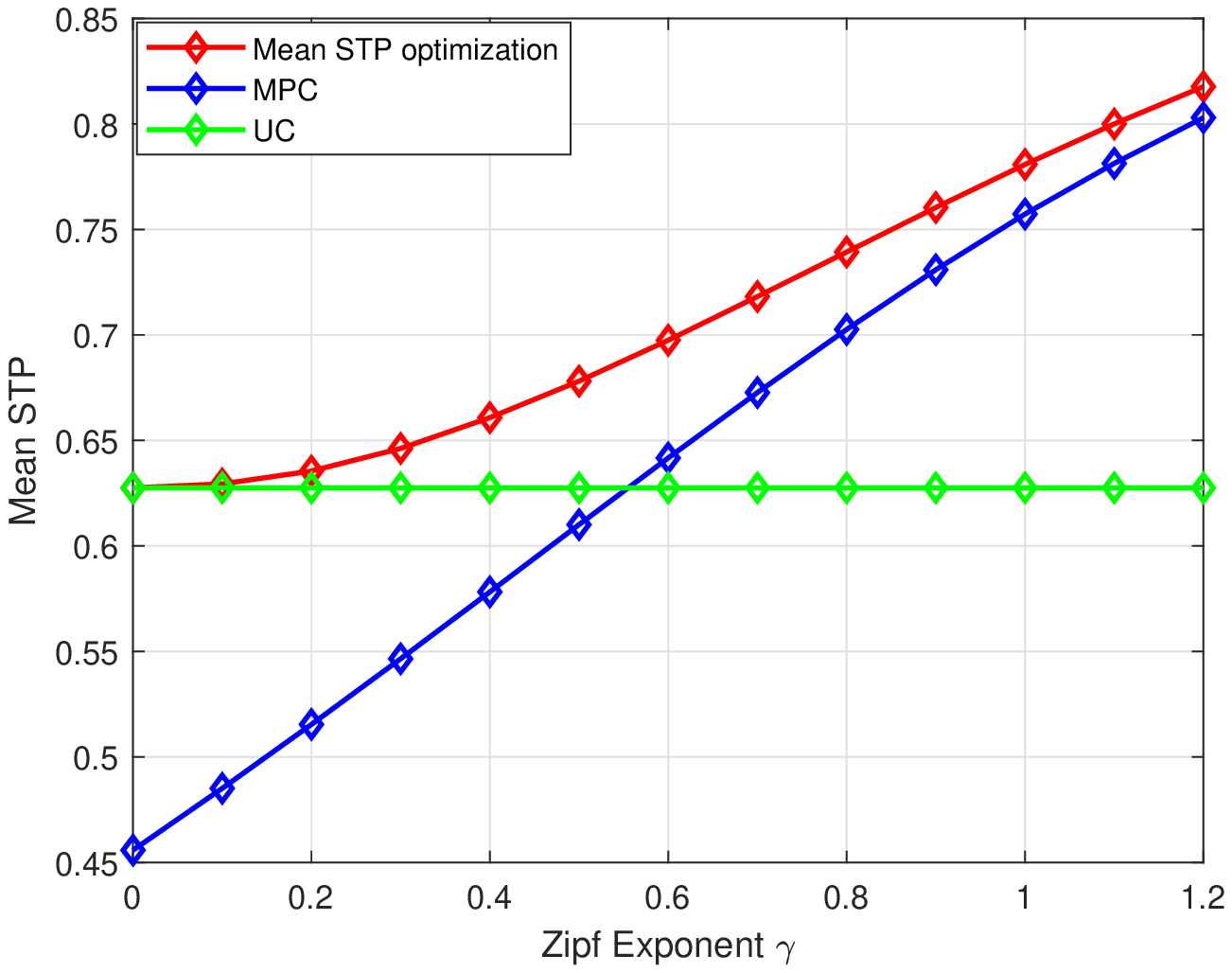}
\caption{Mean STP as functions of Zipf exponent $\gamma$ under different caching strategies.}\label{STP-zipf-exponent}
\end{minipage}%
\hfill
\begin{minipage}[t]{0.5\linewidth}
\centering
\subfigure[Average system transmission delay as functions of backhaul delay $\xi$ for cache size $C=10$.]{
\includegraphics[width=3.5in]{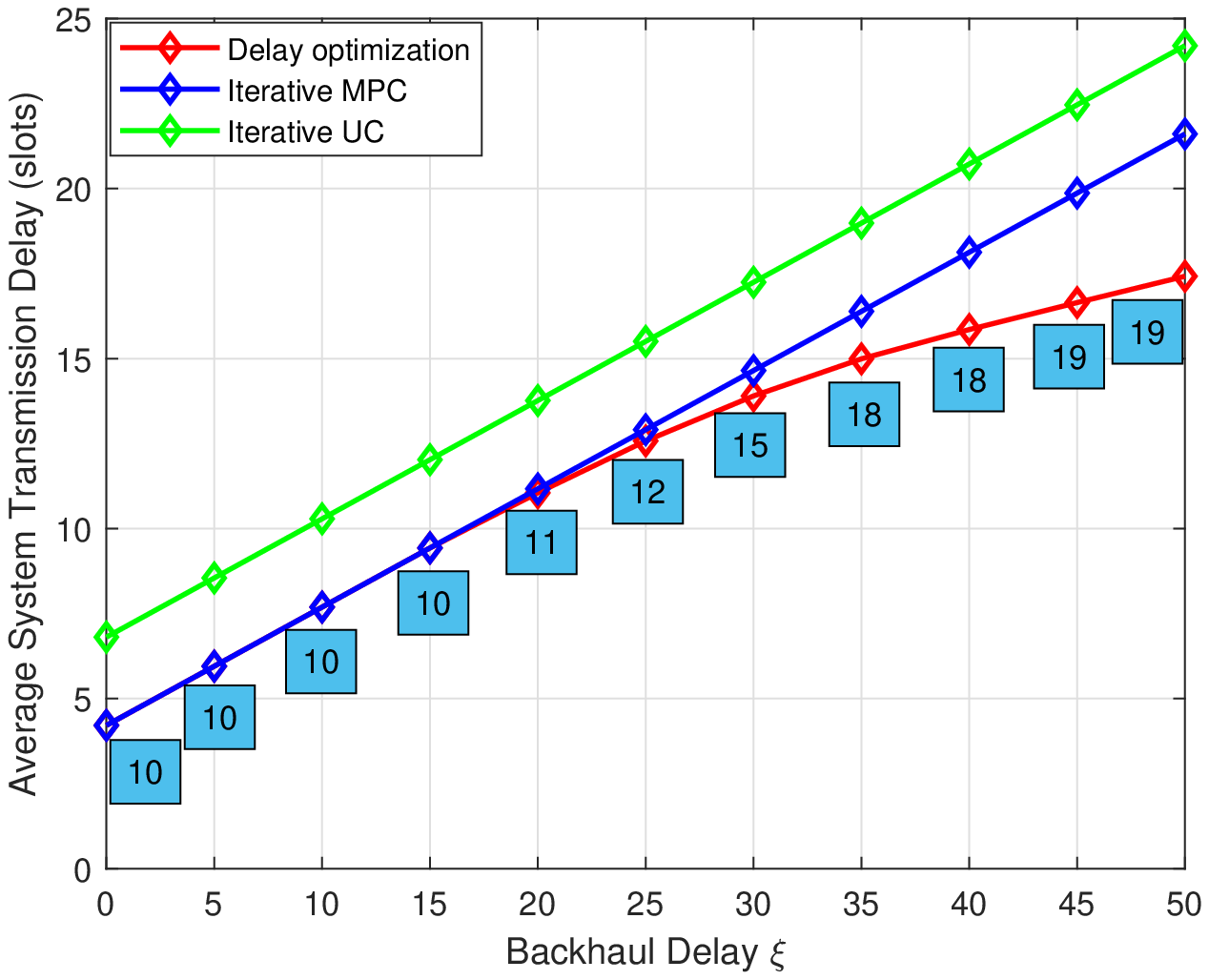}
}
\end{minipage}
\end{figure}

Fig. \ref{STP-zipf-exponent} plots the mean STP as the function of the Zipf exponent $\gamma$, and we can observe that the gap between the proposed method and the MPC decreases with $\gamma$. The reason is that the majority of the user requests are concentrated on fewer files when $\gamma$ increases. Note that the performance of the UC strategy remains invariant with $\gamma$ due to the fact that the caching probability for each file keeps unchanged. These are consistent with the results in \cite{wen}, but our meta distribution based approach in this paper has offered a much more general analysis framework.

\begin{figure}
\subfigure[Average system transmission delay as functions of backhaul delay $\xi$ for cache size $C=15$.]{
  \includegraphics[width=3.5in]{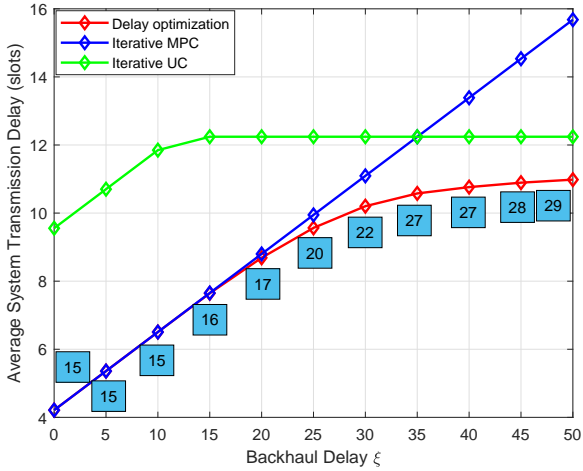}
  }
\subfigure[Average system transmission delay as functions of backhaul delay $\xi$ for cache size $C=20$.]{
  \includegraphics[width=3.5in]{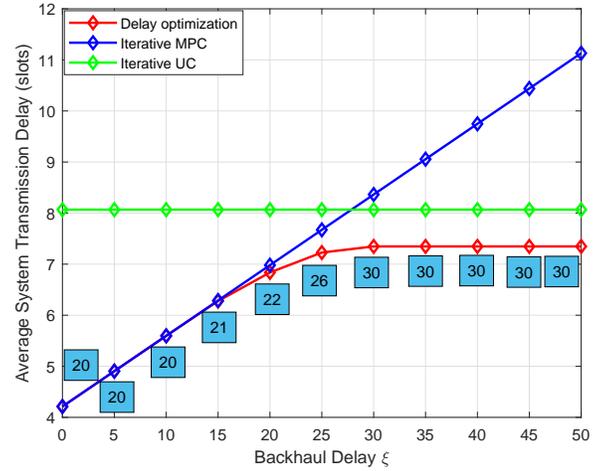}
  }
  \caption{The impact of the backhaul delay on the average system transmission delay under different cache sizes.}\label{service-delay-backhaul-delay-cache-size}

\end{figure}

Fig. \ref{service-delay-backhaul-delay-cache-size} plots the effect of the backhaul delay $\xi$ on the average system transmission delay (i.e., $D_{ai}$ in (\ref{D-ai-equation})) under the proposed algorithm and two other baseline caching strategies, i.e., the iterative MPC and iterative UC strategies. Note that the numbers of files cached in the BSs under different caching strategies are marked in the blue boxes. Both baseline strategies undergo an iterative process to obtain the optimal BS active probability. The differences lie in that (1) for the iterative UC strategy, the caching probability $q_f=C/\min\left\{\left\lceil\frac{C}{q_c}\right\rceil-1,F\right\}$ for $f\in\left[1,\min\left\{\left\lceil\frac{C}{q_c}\right\rceil-1,F\right\}\right]$ and $q_f=0$ for $f\in\left[\min\left\{\left\lceil\frac{C}{q_c}\right\rceil-1,F\right\}+1,F\right]$ in each iteration; and (2) for the iterative MPC strategy, the caching probability $q_f=1$ for $f\in\left[1,C\right]$ and $q_f=0$ for $f\in\left[C+1,F\right]$ in each iteration. From \cite{backhaul-delay} and \cite{backhaul-aware}, the backhaul delay of a file can range from 25\% to 130\% of its local delay. In order to investigate the effect of such delay, the range of backhaul delay is set to be 0--50 slots (given the normal range for local delay in Figs. \ref{local-delay-caching-active} and \ref{local-delay-active-caching}).

From Fig. \ref{service-delay-backhaul-delay-cache-size}, it can be observed that the average system transmission delay increases with the backhaul delay for the proposed algorithm and the iterative MPC under different cache sizes and the proposed optimization algorithm always performs better than both the iterative MPC and iterative UC strategies. When the backhaul delay $\xi$=0, 5 or 10, the number of cached files is equal to the cache size and the MPC strategy performs the same as the proposed method. When the backhaul delay increases, the optimal number of cached files tends to be larger than the cache size. The reason is that the users are less likely to suffer from the severe backhaul delay when more files can be directly obtained from the BSs. Note that when the cache size $C=$10 (i.e., Fig. \ref{service-delay-backhaul-delay-cache-size}(a)), the average system transmission delay for the iterative UC is always larger than the other two strategies. When the cache size $C=$15 (i.e., Fig. \ref{service-delay-backhaul-delay-cache-size}(b)) or 20 (i.e., Fig. \ref{service-delay-backhaul-delay-cache-size}(c)), the performance gap between the iterative UC strategy and the proposed algorithm becomes smaller in the large backhaul delay regime. In addition, the average system transmission delay for the iterative UC strategy becomes lower than that for the iterative MPC strategy in the large backhaul delay regime, indicating that better performance can be achieved by seeking larger file diversity rather than only attempting to cache the most popular files. In summary, it is better to cache only the most popular files when the backhaul delay is small while a larger file diversity will be more beneficial when the backhaul delay is large.

\begin{figure}[htbp]
\begin{minipage}[t]{0.35\linewidth}
\centering
\includegraphics[width=3.5in]{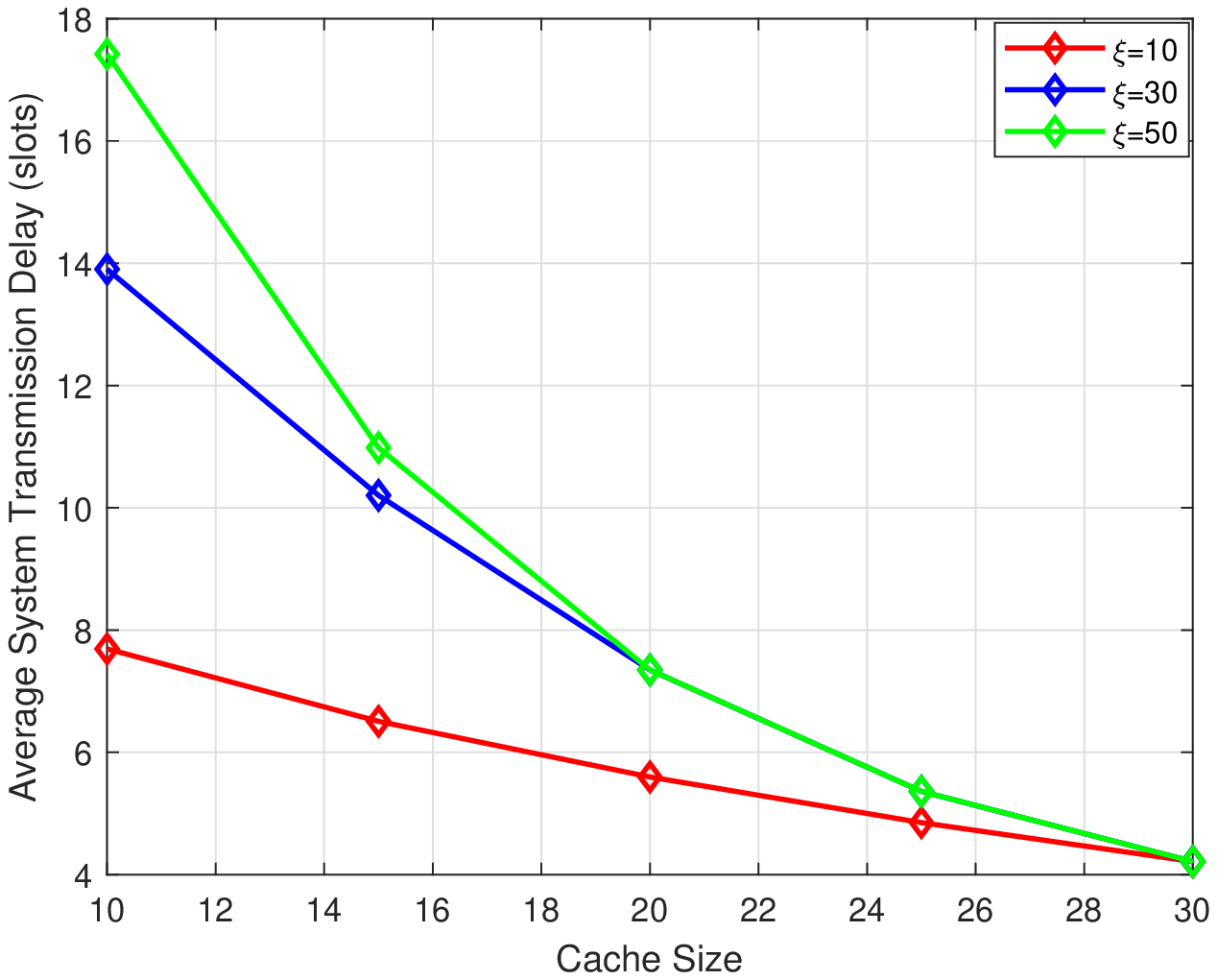}
\caption{Average system transmission delay as functions of cache size $C$ under different backhaul delays.}\label{local-delay-cache-size-backhaul-delay}
\end{minipage}%
\hfill
\begin{minipage}[t]{0.5\linewidth}
\centering
\includegraphics[width=3.5in]{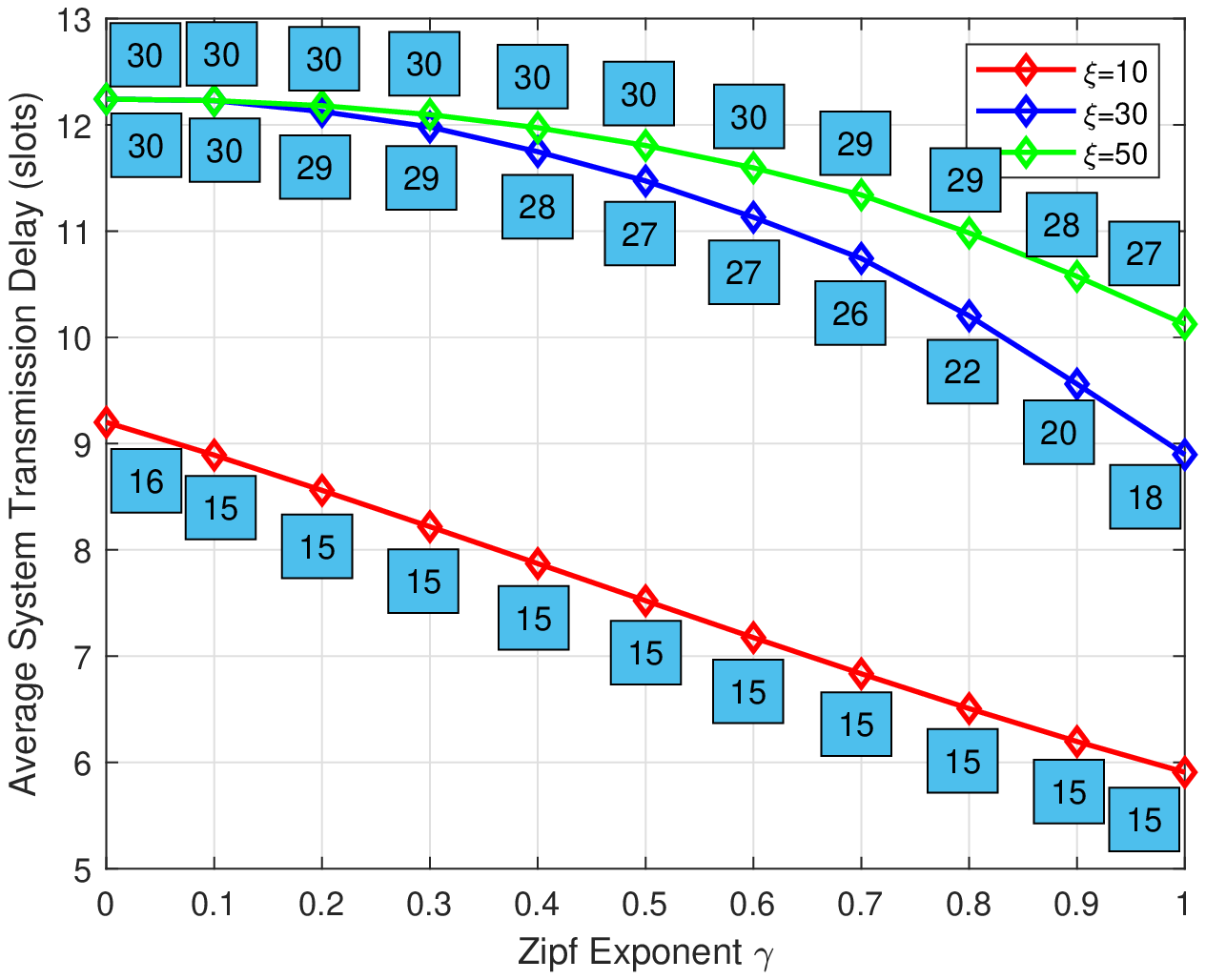}
\caption{Average system transmission delay as functions of Zipf exponent $\gamma$ under different backhaul delays.}\label{local-delay-zipf-exponent-backhaul-delay}
\end{minipage}
\end{figure}

Fig. \ref{local-delay-cache-size-backhaul-delay} illustrates the average system transmission delay as a function of the cache size under different backhaul delays. We can observe that the optimal average system transmission delay decreases with the cache size. This is because the increasing cache size can not only enhance the file diversity but also shorten the distance between the user and the serving BS, leading to the increase of the SIR and the decrease of the mean local delay. Moreover, the average system transmission delay increases with the backhaul delay and the performance gap between the average system transmission delays under different backhaul delays become smaller with the increasing cache size. This can be explained as follows. First, when the backhaul delay increases, a considerable amount of time would be taken by a file for its transmission over the backhaul link when retrieving the file from the core network. Therefore, the backhaul transmission should be avoided as much as possible. Second, when the cache size increases, a larger number of files are cached in the BSs and fewer files need to be retrieved from the core network through the backhaul link. Note that the average system transmission delays become identical when the cache size is 30 since all the files can be cached in the BSs and no file needs to be retrieved from the core network.

Fig. \ref{local-delay-zipf-exponent-backhaul-delay} plots the impact of the Zipf exponent on the average system transmission delay. Similarly to Fig. \ref{service-delay-backhaul-delay-cache-size}, the numbers of files cached in the BSs under different caching strategies are marked in the blue boxes. It can be observed that the average system transmission delay decreases with the Zipf exponent $\gamma$ and fewer files are cached in the BSs when $\gamma$ increases. The reason is that when the Zipf exponent $\gamma$ increases, a larger number of user requests are concentrated on fewer popular files, indicating that the less popular files which may suffer from the severe backhaul delay are less likely to be requested. Hence, the average system transmission delay of the overall network decreases.
\section{Conclusion}
We have provided a fine-grained analysis on the STP and transmission delay of the cache-enabled networks. The moments of the conditional STP, the exact meta distribution and its beta approximation were derived by utilizing stochastic geometry. The closed-form expression of the mean local delay was also derived. We considered the maximization of the STP and the minimization of the mean local delay by optimizing the caching probability and the BS active probability, respectively. For the former, a convex optimization problem was formulated and the optimal caching probability and BS active probability were achieved. For the latter, a non-convex optimization problem was formulated and an iterative algorithm was proposed to obtain the optimal solution. We demonstrated the impact of the backhaul delay on the caching strategy. That is, MPC is optimal when the backhaul delay is relatively low but the network performance benefits from a larger file diversity when the backhaul delay is very large. However, the proposed optimization algorithms always perform better than both MPC and UC. More importantly, the structure of the optimal caching strategy provides useful insights for designing the cache-enabled networks. Extension work for the future include the analysis for the non-PPP BS distributions and examination of the impact of file popularity prediction errors.

\section{Appendices}
\subsection{Proof of Theorem \ref{theorem-moment}}
Assuming that $u_0$ requests File $f$, the STP conditioned on the realization $\Phi$ can be derived as
\begin{equation}
\begin{split}
\mathcal{P}(\theta|\Phi)&\overset{(a)}{=}\beta\mathbbm{P}\left[\text{SIR}_{k,t}>\theta\mid\Phi\right]
=\beta\mathbbm{P}\left[P|h_{i_0}|^2r^{-\alpha}>\theta(I_{t,f}+I_{t,-f})\mid\Phi\right]\\
&\overset{(b)}{=}\beta\mathbbm{E}_{I}\left[\exp\left(-\frac{\theta r^{\alpha}}{P}I_{t,f}\right)
\exp\left(-\frac{\theta r^{\alpha}}{P}I_{t,-f}\right)\right]\\
&\overset{(c)}{=}\beta\mathbbm{E}_{\Phi}\left[\mathcal{L}_{I_{t,f}}\left(\left.\frac{\theta r^{\alpha}}{P}\right|\Phi\right)
\mathcal{L}_{I_{t,-f}}\left(\left.\frac{\theta r^{\alpha}}{P}\right|\Phi\right)\right],
\end{split}
\end{equation}
where (a) follows from the definition of $\mathcal{P}(\theta|\Phi)$, and (b) follows from $h\sim\exp(1)$. In the last step, $\mathcal{L}_{I_{t,f}}\left(\left.\frac{\theta r^{\alpha}}{P}\right|\Phi\right)$ and $\mathcal{L}_{I_{t,-f}}\left(\left.\frac{\theta r^{\alpha}}{P}\right|\Phi\right)$ denote the Laplace transforms of the interferences of the BSs with/without caching File $f$, respectively. Assuming $s=\frac{\theta r^{\alpha}}{P}$, the Laplace transform can be derived as
\begin{equation}\label{laplace-f}
\begin{split}
\mathcal{L}_{I_{t,f}}(s|\Phi)&=\mathbbm{E}_{I_{t,f}}\left[\exp\left(-sI_{t,f}\right)\right]
=\mathbbm{E}_{h_{i}}\left[\exp\left(-s\sum_{i\in\Phi_{f}\backslash i_0}\mathbf{1}(i\in\Phi_{-f}(t))Ph_ix_i^{-\alpha}\right)\right]\\
&=\prod_{i\in\Phi_{t,f}\backslash i_0}\mathbbm{E}_{h_{i}}\left[\beta\exp\left(-sPh_ix_i^{-\alpha}\right)+1-\beta\right]\\
&=\prod_{i\in\Phi_{t,f}\backslash i_0}\left(\frac{\beta}{1+sPx_i^{-\alpha}}+1-\beta\right).
\end{split}
\end{equation}

Similarly, the Laplace transform of the interferences from the BSs not caching File $f$ can be evaluated as
\begin{equation}\label{laplace-no-f}
\begin{split}
\mathcal{L}_{I_{t,f}}(s|\Phi)
=\prod_{i\in\Phi_{t,-f}}\left(\frac{\beta}{1+sPx_i^{-\alpha}}+1-\beta\right).
\end{split}
\end{equation}

Accordingly, the $k$-th moment of the conditional STP can be derived as
\begin{equation}\label{moment-proof}
\begin{split}
M_{k,f}=&\mathbbm{E}_{\Phi}\left[\mathbbm{P}^k\left[\text{SIR}>\theta\mid\Phi\right]\right]
=\beta\mathbbm{E}_{\Phi}\left[\mathcal{L}_{I_{t,f}}\left(\left.\frac{\theta r^{\alpha}}{P}\right|\Phi\right)^k
\mathcal{L}_{I_{t,-f}}\left(\left.\frac{\theta r^{\alpha}}{P}\right|\Phi\right)^k\right].
\end{split}
\end{equation}

Assuming $s=\frac{\theta r^{\alpha}}{P}$, $\mathcal{L}_{I_{t,f}}\left(\left.s\right|\Phi\right)^k$ can be derived as
\begin{equation}
\begin{split}
&\mathcal{L}_{I_{t,f}}\left(\left.s\right|\Phi\right)^k
=\prod_{i\in\Phi_{t,f}\backslash i_0}\left(\frac{\beta}{1+sPx_i^{-\alpha}}+1-\beta\right)^k\\
&\overset{(a)}{=}\exp\left(-2\pi\lambda q_f\int_{r}^{\infty}
\left(1-\left(\frac{\beta}{1+\theta r^{\alpha}x^{-\alpha}}+1-\beta\right)^k\right)x\text{d}x\right)\\
&=\exp\left(-2\pi\lambda q_f\int_{r}^{\infty}
\sum_{n=1}^{\infty}\binom{k}{n}(-1)^{n+1}\left(\frac{\beta\theta r^{\alpha}x^{-\alpha}}{1+\theta r^{\alpha}x^{-\alpha}}\right)^nx\text{d}x\right)\\
&=\exp\left(-\pi\delta\lambda q_f\sum_{n=1}^{\infty}\binom{k}{n}(-1)^{n+1}
\frac{(\beta\theta)^nr^2}{(n-\delta)}
F\left(n,n-\delta,n-\delta+1,-\theta\right)\right).
\end{split}
\end{equation}

Similarly, $\mathcal{L}_{I_{t,-f}}\left(\left.s\right|\Phi\right)^k$ can be derived as
\begin{equation}
\begin{split}
\mathcal{L}_{I_{t,-f}}\left(\left.s\right|\Phi\right)^k
=\exp\left(-\pi\delta\lambda (1-q_f)\sum_{n=1}^{\infty}\binom{k}{n}(-1)^{n+1}
\beta^n\theta^{\delta}r^2\text{B}(\delta,n-\delta)\right).
\end{split}
\end{equation}

According to \cite{distance-pdf}, the probability density function (PDF) of the distance between $u_0$ and the serving BS conditioned on File $f$ being requested is given by
\begin{equation}\label{pdf}
f_R(r)=2\pi\lambda q_{f}r\exp(-\pi\lambda q_fr^2)
\end{equation}

The $k$-th moment of the conditional STP in (\ref{moment-proof}) can then be written further as
\begin{equation}
\begin{split}
M_{k,f}&=\int_{0}^{\infty}2\pi\lambda q_fr\exp\left(-\pi\lambda q_fr^2\right)
\mathcal{L}_{I_{t,f}}\left(\left.\frac{\theta r^{\alpha}}{P}\right|\Phi\right)^k
\mathcal{L}_{I_{t,-f}}\left(\left.\frac{\theta r^{\alpha}}{P}\right|\Phi\right)^k\text{d}r\\
&=q_f\left(q_f+\delta(1-q_f)\sum_{n=1}^{\infty}\binom{k}{n}(-1)^{n+1}\beta^n\theta^{\delta}\text{B}(n,n-\delta)\right.\\
&\left.+\delta q_f\sum_{n=1}^{\infty}\binom{b}{n}(-1)^{n+1}\frac{(\beta\theta)^n}{n-\delta}F\left(n,n-\delta,n-\delta+1,-\theta\right)\right)^{-1},
\end{split}
\end{equation}
which is Theorem \ref{theorem-moment} (i.e., Eq. (\ref{moment-equation})), and $F(\cdot)$ and $\text{B}(\cdot)$ are defined therein.

\subsection{Proof of Theorem \ref{theorem-delay}}
First, the mean local delay can be derived as
\begin{equation}
\begin{split}
M_{-1,f}&=\mathbbm{E}_{\Phi}\left[\mathbbm{P}^{-1}\left[\text{SIR}>\theta\mid\Phi\right]\right]
=\beta\mathbbm{P}\left[P|h_{i_0}|^2r^{-\alpha}>\theta(I_{t,f}+I_{t,-f})\mid\Phi\right]^{-1}\\
&=\beta\mathbbm{E}_{\Phi}\left[\frac{1}{\mathcal{L}_{I_{t,f}}\left(\left.\frac{\theta r^{\alpha}}{P}\right|\Phi\right)}
\frac{1}{\mathcal{L}_{I_{t,-f}}\left(\left.\frac{\theta r^{\alpha}}{P}\right|\Phi\right)}\right].
\end{split}
\end{equation}

We then derive the expression of $\frac{1}{\mathcal{L}_{I_{t,f}}(s)}$ as follows.
\begin{equation}
\begin{split}
\frac{1}{\mathcal{L}_{I_{t,f}}(s)}
&\overset{(a)}{=}\exp\left(-2\pi\lambda\int_{r}^{\infty}\left(1-\left(\frac{\beta}{1+sPx^{-\alpha}}+1-\beta\right)^{-1}\right)x\text{d}x\right)\\
&\overset{(b)}{=}\exp\left(2\pi\lambda\int_{r}^{\infty}\left(\frac{\beta\theta r^{\alpha}x^{-\alpha}}{1+(1-\beta)\theta r^{\alpha}x^{-\alpha}}\right)x\text{d}x\right)\\
&=\exp\left(\pi\delta\lambda q_f\beta\theta r^{\alpha}\int_{r^{\alpha}}^{\infty}
\frac{u^{\delta-1}}{u+(1-\beta)\theta r^{\alpha}}\text{d}u\right)\\
&=\exp\left(\pi\delta\lambda q_f\frac{\beta\theta r^2}{1-\delta}F\left(1,1-\delta,2-\delta,-(1-\beta)\theta\right)\right),
\end{split}
\end{equation}
where (a) is obtained by utilizing the probability generating functional (PGFL) of the PPP, and (b) follows from the binomial theorem.

Similarly, the derivation of $\frac{1}{\mathcal{L}_{I_{t,-f}}(s)}$ can be obtained:
\begin{equation}
\begin{split}
\frac{1}{\mathcal{L}_{I_{t,-f}}(s)}&=\exp\left(\pi\delta\lambda(1-q_f)\beta\theta r^{\alpha}\int_{0}^{\infty}
\frac{u^{\delta-1}}{u+(1-\beta)\theta r^{\alpha}}\text{d}u\right)\\
&=\exp\left(\pi\delta\lambda(1-q_f)\beta\theta^{\delta}r^2\text{B}(\delta,1-\delta)\right).
\end{split}
\end{equation}

The local delay for File $f$ can then be derived by averaging over $r$ as
\begin{equation}
\begin{split}
&M_{-1,f}=\int_{0}^{\infty}\frac{2\pi\lambda q_fr}{\mathcal{L}_{I_{t,f}}(\left.\frac{\theta r^{\alpha}}{P}\right|\Phi)\mathcal{L}_{I_{t,-f}}(\left.\frac{\theta r^{\alpha}}{P}\right|\Phi)}\exp\left(-\pi\lambda q_fr^2\right)\text{d}r\\
\overset{(a)}{=}&q_f\left(\beta\left(q_f-\delta(1-q_f)(1-\beta)^{\delta-1}\beta\theta^{\delta}\text{B}(\delta,1-\delta)
-\frac{q_f\delta\beta\theta }{1-\delta}F\left(1,1-\delta,2-\delta,-(1-\beta)\theta\right)\right)\right)^{-1},
\end{split}
\end{equation}
where (a) follows from the fact that $\int_{0}^{\infty}2re^{-Ar^2}=1/A$. This completes the proof for Theorem \ref{theorem-delay}.

\subsection{Proof of Lemma \ref{lemma-laplace}}
By establishing the Lagrange function of Problem 1, we can obtain
\begin{equation}
\begin{split}
L(\mathbf{q},\mathbf{\rho},\mathbf{\upsilon},\eta)=&D+\sum_{f=1}^{F}\rho_f(q_f-q_c)
+\sum_{f=1}^{F}\upsilon_f(1-q_f)+\eta
\left(C-\sum_{f=1}^{F}q_f\right),
\end{split}
\end{equation}
where $\rho_f$,$\upsilon_f$ and $\eta$ are the Lagrange multipliers. $\rho_f$ and $\upsilon_f$ are associated with (1), and $\eta$ is associated with (2). Note that $\mathbf{\rho}\triangleq(\rho_f)_{f\in\mathcal{F}}$ and $\mathbf{\upsilon}\triangleq(\upsilon_f)_{f\in\mathcal{F}}$. Thus, we have
\begin{equation}\label{objective-function}
\frac{\partial L(\mathbf{q},\mathbf{\rho},\mathbf{\upsilon},\eta)}{\partial q_f}=\frac{-C_1}{\beta\left(C_3q_f-C_1\right)^2}+\rho_f-\upsilon_f-\eta.
\end{equation}

The KKT conditions can be written as
\begin{equation}\label{constraint-1}
\frac{\partial L(\mathbf{q}^{*},\mathbf{\rho},\mathbf{\upsilon},\eta)}{\partial q_f}=0,\,\forall f\in\mathcal{F},
\end{equation}
\begin{equation}\label{complementary-slackness}
\rho_f(q_f^{*}-q_c)=0,\, \upsilon_f(1-q_f^{*})=0,\, \eta(C-\sum_{f=1}^{F}q_f^{*})=0,\, \forall f\in\mathcal{F}
\end{equation}
\begin{equation}
\sum_{f=1}^{F}q_f^{*}=C,\, 0\leq q_f^{*}\leq 1,\,\forall f\in\mathcal{F},
\end{equation}
\begin{equation}\label{dual-constraint}
\rho_f\geq 0, \upsilon_f\geq 0,\,\forall f\in\mathcal{F},
\end{equation}
where (\ref{complementary-slackness}) is the complementary slackness and (\ref{complementary-slackness}) is the dual constraint.

According to (\ref{objective-function}) and (\ref{constraint-1}), we have
\begin{equation}
\eta=\frac{-C_1}{\beta\left(C_3q_f-C_1\right)^2}+\rho_f-\upsilon_f,\,\forall f\in\mathcal{F}.
\end{equation}

Next, we analyze the optimal solution by considering three cases as follows.
\begin{enumerate}
\item If $q_f^*=q_c$, then $\rho_f\geq 0$, $\upsilon_f=0$ and $\eta=\frac{-C_1}{\beta\left(C_3q_f-C_1\right)^2}+\rho_f$, implying that $\eta\geq\frac{-C_1}{\beta\left(C_3q_f-C_1\right)^2}$.
\item If $q_f^*=1$, then $\rho_f=0$, $\upsilon_f\geq 0$ and $\eta=\frac{-C_1}{\beta\left(C_3q_f-C_1\right)^2}-\upsilon_f$, implying that $\eta\leq\frac{-C_1}{\beta\left(C_3q_f-C_1\right)^2}$.
\item If $q_c<q_f^*<1$, then $\rho_f=0$, $\upsilon_f=0$ and $\eta=\frac{-C_1}{\beta\left(C_3q_f-C_1\right)^2}$.
\end{enumerate}

This completes the proof of Lemma \ref{lemma-laplace}.

\end{document}